\documentclass[sigconf]{acmart}
\usepackage{graphicx}
\usepackage{amsmath,amsfonts}
\usepackage{amsthm}
\usepackage{booktabs}
\usepackage{multirow}
\usepackage{bm}
\usepackage{subfigure}
\usepackage{xcolor}
\usepackage[linesnumbered,ruled,noend]{algorithm2e} 
\usepackage{algorithmic}
\usepackage{siunitx}
\usepackage[switch]{lineno}
\usepackage{paralist}
\usepackage{chngcntr}
\usepackage{hyperref}
\usepackage{wrapfig}
\usepackage{adjustbox}
\usepackage[skip=4pt]{caption}
\AtBeginDocument{%
  \providecommand\BibTeX{{%
    \normalfont B\kern-0.5em{\scshape i\kern-0.25em b}\kern-0.8em\TeX}}}

\copyrightyear{2021}
\acmYear{2021}
\setcopyright{acmcopyright}
\acmConference[CIKM '21] {Proceedings of the 30th ACM International Conference on Information and Knowledge Management}{November 1--5, 2021}{Virtual Event, Australia.}
\acmBooktitle{Proceedings of the 30th ACM Int'l Conf. on Information and Knowledge Management (CIKM '21), November 1--5, 2021, Virtual Event, Australia}
\acmPrice{15.00}
\acmDOI{10.1145/3459637.3482449}
\acmISBN{978-1-4503-8446-9/21/11}

\begin{document}

\title{LT-OCF: Learnable-Time ODE-based Collaborative Filtering}
\author{Jeongwhan Choi, Jinsung Jeon, Noseong Park}
\email{{jeongwhan.choi, jjsjjs0902, noseong}@yonsei.ac.kr}
\affiliation{%
  \institution{Yonsei University}
  \city{Seoul}
  \country{South Korea}
}

\renewcommand{\shortauthors}{Choi, et al.}

\begin{abstract}
Collaborative filtering (CF) is a long-standing problem of recommender systems. Many novel methods have been proposed, ranging from classical matrix factorization to recent graph convolutional network-based approaches. After recent fierce debates, researchers started to focus on \emph{linear} graph convolutional networks (GCNs) with a \emph{layer combination}, which show state-of-the-art accuracy in many datasets. In this work, we extend them based on neural ordinary differential equations (NODEs), because the linear GCN concept can be interpreted as a differential equation, and present the method of Learnable-Time ODE-based Collaborative Filtering (LT-OCF). The main novelty in our method is that after redesigning linear GCNs on top of the NODE regime, i) we learn the optimal architecture rather than relying on manually designed ones, ii) we learn smooth ODE solutions that are considered suitable for CF, and iii) we test with various ODE solvers that internally build a diverse set of neural network connections. We also present a novel training method specialized to our method. In our experiments with three benchmark datasets, our method consistently outperforms existing methods in terms of various evaluation metrics. One more important discovery is that our best accuracy was achieved by dense connections.
\end{abstract}




\maketitle

\section{Introduction}

\begin{figure}[t]
\centering
\subfigure[The architecture of LightGCN. The outer loop created by a series of additions, denoted $\oplus$, is called \emph{layer combination}.]{\includegraphics[width=1\columnwidth]{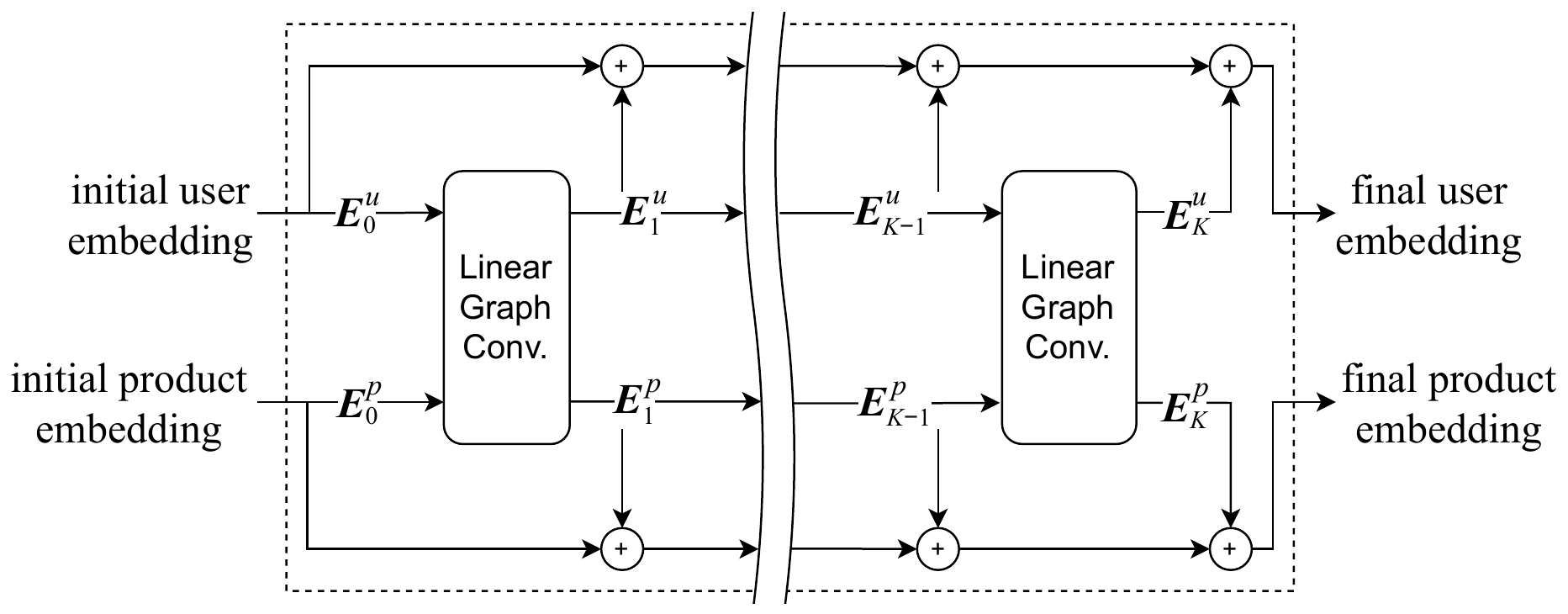}}
\subfigure[Our proposed LT-OCF. The detailed diagram of the ODE layer is in Fig.~\ref{fig:dual}.]{\includegraphics[width=1\columnwidth]{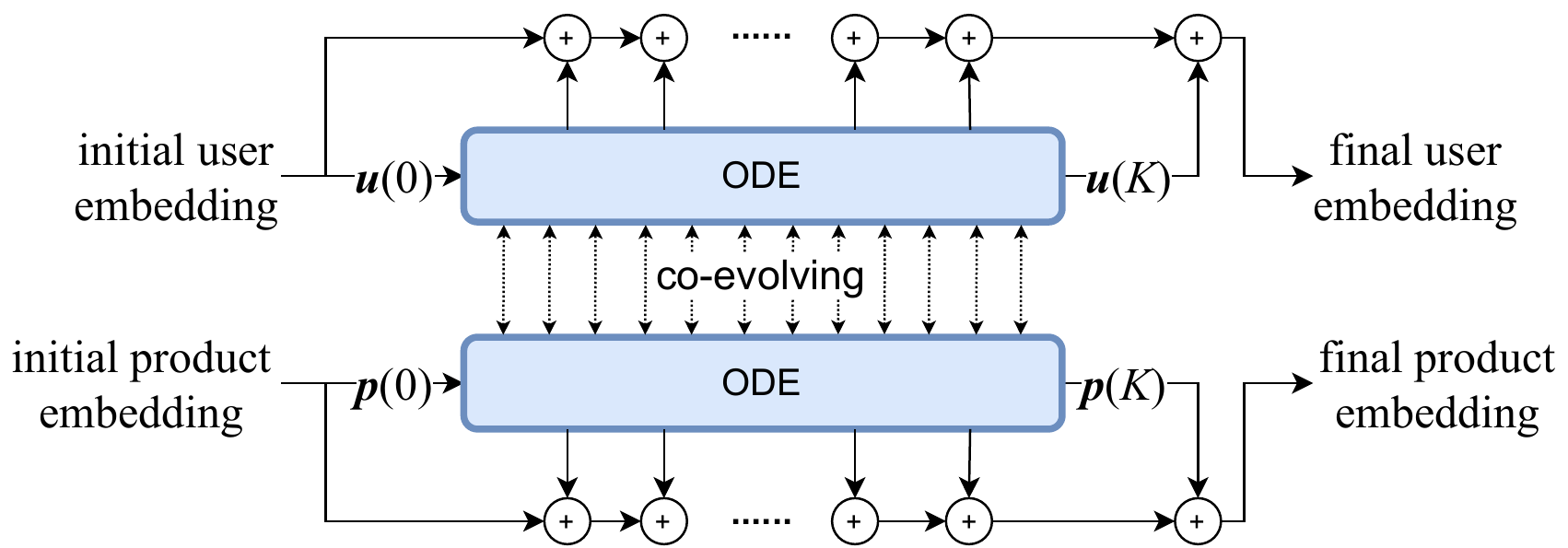}}
\caption{(a) The linear architecture of LightGCN with a layer combination, and (b) Our proposed LT-OCF. In our design, two ODEs for user and product embeddings co-evolve over time and influence each other.}\label{fig:archi}
\end{figure}

Collaborative filtering (CF), which is to predict users' preferences from patterns, is a long-standing research problem in the field of recommender systems~\cite{8506344,10.1145/3285029,Koren2010,Y.Wu2016,He2017, 5197422,Wang2014,Ebesu2018,Adomavicius2015,Yang2016}. It is common to learn user and product embedding vectors and calculate their dot-products for recommendation. Matrix factorization is one such approach, which is well-known in the recommender system community~\cite{5197422}. There have been proposed several other enhancements as well~\cite{10.1145/1401890.1401944,8352808}. Recently researchers started to focus on graph convolutional networks (GCNs) for the purpose of CF~\cite{10.1145/3331184.3331267,10.1145/3397271.3401063,Chen_Wu_Hong_Zhang_Wang_2020}. GCNs had been proposed to process not only CF-related graphs but also other general graphs. GCNs are broadly categorized into the following two types: spectral GCNs~\cite{Bruna2014,Defferrard2016,kipf2017semi,Wu2019,Xu2019} and spatial GCNs~\cite{Atwood2015,Gilmer2017,Hamilton2017,Velickovic2018,Gao2018}. GCNs for CF fall into the first category due to its appropriateness for CF~\cite{10.1145/3397271.3401063,Chen_Lin_Li_Li_Zhou_Sun_2020}.

However, there have been fierce debates about what is the optimal GCN architecture for CF. During its early phase, researchers utilized non-linear activations, such as ReLU, because they showed good accuracy in many machine learning tasks, e.g., classification, regression, and so on~\cite{DBLP:journals/corr/BergKW17,DBLP:conf/kdd/YingHCEHL18,DBLP:conf/mm/WeiWN0HC19,DBLP:conf/www/WangZXLG19,10.1145/3331184.3331267}. Surprisingly, however, it was recently reported that a \emph{linear} GCN architecture with a \emph{layer combination}, called LightGCN, works better than other non-linear GCNs for CF~\cite{10.1145/3331184.3331267,Chen_Wu_Hong_Zhang_Wang_2020,10.1145/3397271.3401063}. Unlike other general graphs, user-product interaction bipartite graphs are frequently sparse and provide little information because they mostly do not include node/edge features. In~\cite{10.1145/3397271.3401063}, it was noted that, for the same reason, non-linear GCNs are quickly overfitted to training data and do not work well in general for CF.

Owing to the discovery, we propose the method of Learnable-Time ODE-based Collaborative Filtering (LT-OCF) in this paper. We redesign the linear GCN with the layer combination on top of the concept of the neural ordinary differential equations (NODEs) because linear GCNs, including LightGCN, can be theoretically interpreted as differential equations, i.e., heat equations (see Section~\ref{sec:thermal}). For instance, the main linear propagation layer of LightGCN is exactly the same as Newton's law of cooling.

Neural ordinary differential equations (NODEs) are to learn implicit differential equations from data. NODEs calculate $\bm{h}(t_1) = \bm{h}(t_0) + \int_{t_0}^{t_1}f(\bm{h}(t),t;\bm{\theta}_f)dt$, where $f$ is a neural network parameterized by $\bm{\theta}_f$ that approximates $\frac{d\bm{h}(t)}{dt}$, to derive $\bm{h}(t_1)$ from $\bm{h}(t_0)$, where $t_1 > t_0$. We note that $\bm{\theta}_f$ is trained from data --- in other words, $\frac{d\bm{h}(t)}{dt}$ is trained from data. The variable $t$ is called as \emph{time variable}, which represents the layer concept of neural networks. Note that $t$ is a non-negative integer in conventional neural networks whereas it can be any arbitrary non-negative real number in NODEs. In this regard, NODEs are considered as \emph{continuous} generalizations of neural networks.

Various ODE solvers can solve the integral problem, and it was also known that they can generalize various neural network architectures~\cite{NIPS2018_7892}. For instance, the general form of the residual connection can be written as $\bm{h}(t+1) = \bm{h}(t) + f(\bm{h}(t);\bm{\theta})$, which is identical to the explicit Euler method to solve ODE problems. It is also known that the fourth-order Runge--Kutta (RK4) ODE solver is similar to dense convolutional networks and fractal neural networks~\cite{pmlr-v80-lu18d}.

The reason of our specific design choice to adopt NODEs for CF is threefold. In our proposed LT-OCF, firstly, $t$ is not only \emph{continuous} but also \emph{trainable} because we interpret linear GCNs as continuous-time differential equations. Therefore, we can learn the optimal layer combination construction rather than relying on a manually configured one. Let $\bm{E}^u_i \in \mathbb{R}^{N \times D}$ and $\bm{E}^p_i \in \mathbb{R}^{M \times D}$, where $N$ is the number of users, $M$ is the number of products, and $D$ is the dimensionality of embedding space, be the user and product embeddings at layer $i$, respectively. In recent GCN-based CF methods~\cite{Chen_Wu_Hong_Zhang_Wang_2020,10.1145/3397271.3401063}, for instance, the final user embeddings are calculated, owing to the \emph{layer combination} technique, by $w_0 \bm{E}^u_0 + w_1\bm{E}^u_1 + w_2\bm{E}^u_2 + \cdots + w_K\bm{E}^u_K$, where $w_i$ is a coefficient and $K$ is the number of layers. On the other hand, one contribution of LT-OCF is to calculate $w_0\bm{u}(0) + w_1\bm{u}(t_1) + w_2\bm{u}(t_2) + \cdots + w_K\bm{u}(t_K)$, where $\bm{u}(k)$ corresponds to $\bm{E}^u_k$, $t_i$ is trainable for all $i$, and $t_i < t_j$ if $i < j$.

Secondly, NODEs learn homeomorphic functions which we consider suitable for CF --- see our discussion after Proposition~\ref{p:hom}. In the recent linear GCN architecture of CF~\cite{10.1145/3397271.3401063}, in addition, user/product embedding is simply a weighted sum of neighbors' embeddings, which can be solely written as matrix multiplications. We also use only matrix multiplications for defining our ODE formulation and matrix multiplication is an analytic operator. The Cauchy--Kowalevski theorem states, in such a case, that the optimal solution of $\bm{h}(t)$ always exists and is unique~\cite{10.2307/j.ctvzsmfgn}. Therefore, our ODE-based CF is a well-posed problem (see Section~\ref{sec:well}).

After formulating CF as an ODE problem, thirdly, we test with various ODE solvers that internally create a rich set of neural network connections. For instance, residual connections are the same as the explicit Euler method, dense connections are the same as RK4, and so on. We can test with various connections.

The architecture of LT-OCF is shown in Fig.~\ref{fig:archi} (b). To enable the proposed concept, we define dual co-evolving ODEs, whose detailed diagram is in Fig.~\ref{fig:dual}. There exists an ODE for each of the user and product embeddings. However, they influence each other and co-evolve over time in our architecture. We also propose a novel training method to train LT-OCF because we have to train dual co-evolving ODEs with their time points $\{t_1, t_2, \cdots\}$.

We use three CF datasets, Gowalla, Yelp2018, and Amazon-Book, and compare LT-OCF with state-of-the-art methods, such as NGCF~\cite{10.1145/3331184.3331267}, LightGCN~\cite{10.1145/3397271.3401063}, and so forth, to name a few. Our method consistently outperforms all those methods in all cases. The biggest enhancement is made for Amazon-Book, i.e., a recall of 0.0411 by LightGCN vs. 0.0442 by LT-OCF and an NDCG of 0.0315 by LightGCN vs. 0.0341 by LT-OCF. We also show that i) our method can be trained faster than LightGCN and ii) dense connections are better than linear connections for CF. To our knowledge, we are the first who reports that dense connections outperform linear connections in CF. Our contributions can be summarized as follows:
\begin{enumerate}
    \item We revisit state-of-the-art linear GCNs and propose the method of Learnable-Time ODE-based Collaborative Filtering (LT-OCF) based on NODEs.
    \item In LT-OCF, we learn the optimal layer combination rather than relying on manually designed architectures.
    \item We reveal that dense connections are better than linear connections for CF (see Section~\ref{sec:dis}). To our knowledge, we first report this observation.
    \item We show that our formulation is theoretically well-posed, i.e., its solution always exists and is unique (see Section~\ref{sec:tra}).
    \item LT-OCF consistently outperforms all existing methods in three benchmark datasets.
\end{enumerate}

\section{Preliminaries \& Related Work}

We introduce our literature survey and preliminary knowledge to understand our work.

\begin{figure}
    \centering
    \includegraphics[width=0.45\columnwidth]{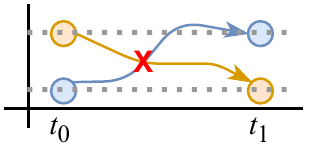}
    \caption{The locations of the yellow and blue points are inverted by the mapping from $t_0$ to $t_1$. NODEs are not able to learn the yellow and blue trajectories at the same time, which cross each other, because their topology (i.e., their relative positions) cannot be changed after a mapping in NODEs, i.e., homeomorphic mapping.}
    \label{fig:cross}
\end{figure}

\subsection{Neural Ordinary Differential Equations (NODEs)} NODEs calculate $\bm{h}(t_{i+1})$ from $\bm{h}(t_i)$ by solving the following Riemann integral problem~\cite{NIPS2018_7892}:
\begin{linenomath*}\begin{align}\label{eq:ode}
    \bm{h}(t_{i+1}) = \bm{h}(t_i) + \int_{t_i}^{t_{i+1}}f(\bm{h}(t),t;\bm{\theta}_f)dt,
\end{align}\end{linenomath*}where the ODE function $f$ parameterized by $\bm{\theta}_f$ is a neural network to approximate the time-derivative of $\bm{h}$, i.e., $\dot{\bm{h}} \stackrel{\text{def}}{=} \frac{d \bm{h}(t)}{d t}$. To solve the problem, we typically rely on existing ODE solvers, e.g., the explicit Euler method, the Dormand--Prince (DOPRI) method, and so forth~\cite{DORMAND198019}.

Let $\phi_t : \mathbb{R}^{\dim(\bm{h}(t_0))} \rightarrow \mathbb{R}^{\dim(\bm{h}(t_1))}$ be a mapping function from $t_0$ to $t_1$ created by Eq.~\eqref{eq:ode}. It is well-known that $\phi_t$ becomes a homeomorphic mapping: $\phi_t$ is bijective and continuous, and $\phi_t^{-1}$ is also continuous for $t \in [0,T]$, where $T$ is the last value in the time domain~\cite{NIPS2019_8577,massaroli2020dissecting}. From this characteristic, the following proposition can be easily proved:
\begin{proposition}\label{p:hom}
The topology of the input space of $\phi_t$ is maintained in the output space, and as a result, the trajectories crossing each other cannot be learned by NODEs, e.g., Fig.~\ref{fig:cross}.
\end{proposition}

 While maintaining the topology, NODEs can perform downstream tasks and it was demonstrated that it actually enhances the robustness to adversarial attacks and out-of-distribution inputs~\cite{yan2020robustness}. We conjecture that this characteristic is also suitable for learning reliable user/product representations, i.e., embeddings, when there is no abundant information. As mentioned earlier, CF typically includes only user-product interactions without additional user/product features. LightGCN~\cite{10.1145/3397271.3401063} showed that, in such a case, linear GCNs with zero non-linearity, which are known to be \emph{smooth}~\cite{Chen_Lin_Li_Li_Zhou_Sun_2020}, are appropriate. We conjecture that NODEs that learn smooth (homeomorphic) functions are also suitable for CF for the same reason. There are several other similar cases where NODEs work well~\cite{kim2021oct,jhin2021acenode,lightmove}.


Instead of the backpropagation method, the adjoint sensitivity method can be adopted and its efficiency and theoretical correctness were already well proved~\cite{NIPS2018_7892}. After letting $\bm{a}_{\bm{h}}(t) = \frac{d L}{d \bm{h}(t)}$ for a task-specific loss $L$, it calculates the gradient of loss w.r.t model parameters with another reverse-mode integral as follows:\begin{align*}\nabla_{\bm{\theta}_f} L = \frac{d L}{d \bm{\theta}_f} = -\int_{t_m}^{t_0} \bm{a}_{\bm{h}}(t)^{\mathtt{T}} \frac{\partial f(\bm{h}(t), t;\bm{\theta}_f)}{\partial \bm{\theta}_f} dt.\end{align*}

We customize the aforementioned adjoint sensitivity method to design our own training algorithm. In our framework, we learn both user/product embeddings and time points $\{t_1, t_2, \cdots\}$ to construct a layer combination using the modified method.

It is known that NODEs have a couple of advantages. First, NODEs can sometimes significantly reduce the required number of parameters when building neural networks~\cite{2019arXiv191010470P}. Second, NODEs enable us to interpret the time variable $t$ as continuous, which is discrete in conventional neural networks~\cite{NIPS2018_7892}. We fully enjoy the second advantage while designing our method.

\begin{figure}[!t]
\centering
\includegraphics[width=1\columnwidth]{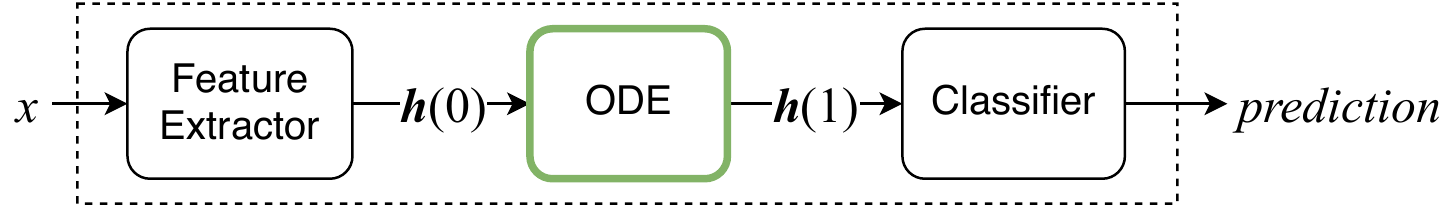}
\caption{The typical architecture of NODEs}\label{fig:archi2}
\end{figure}

Fig.~\ref{fig:archi2} shows the typical architecture of NODEs which we took from~\cite{NIPS2019_8577} --- we assume a downstream classification task in this figure. There is a feature extraction layer which provides $\bm{h}(0)$, and $\bm{h}(1)$ is calculated by the method described above. After that, there is a classification layer. In our case, however, we use the architecture in Fig.~\ref{fig:archi} (b), which has dual co-evolving ODEs only, because our task is not classification but CF.

\subsection{Residual/Dense Connections and ODE Solvers} Many researchers discuss about the analogy between residual/dense connections and ODE solvers. ODE solvers discretize time variable $t$ and convert an integral into many steps of additions. For instance, the explicit Euler method can be written as follows in a step:
\begin{linenomath*}\begin{align}\label{eq:euler}
\bm{h}(t + s) = \bm{h}(t) + s \cdot f(\bm{h}(t), t;\bm{\theta}_f),
\end{align}\end{linenomath*}where $s$, which is usually smaller than 1, is a configured step size of the Euler method. Note that this equation is identical to a residual connection when $s=1$.

Other ODE solvers use more complicated methods to update $\bm{h}(t + s)$ from $\bm{h}(t)$. For instance, the fourth-order Runge--Kutta (RK4) method uses the following method:
\begin{linenomath*}\begin{align}\label{eq:rk4}
\bm{h}(t + s) = \bm{h}(t) + \frac{s}{6}\Big(f_1 + 2f_2 + 2f_3 + f_4\Big),
\end{align}\end{linenomath*}where $f_1 = f(\bm{h}(t), t;\bm{\theta}_f)$, $f_2 = f(\bm{h}(t) + \frac{s}{2}f_1, t+\frac{s}{2};\bm{\theta}_f)$, $f_3 = f(\bm{h}(t) + \frac{s}{2}f_2, t+\frac{s}{2};\bm{\theta}_f)$, and $f_4 = f(\bm{h}(t)+sf_3, t+s;\bm{\theta}_f)$.

It is also known that dense convolutional networks (DenseNets~\cite{zhu2019convolutional}) and fractal neural networks (FractalNet~\cite{Larsson2017FractalNetUN}) are similar to RK4 (as so are residual networks to the explicit Euler method)~\cite{pmlr-v80-lu18d}. For simplicity but without loss of generality, however, we use the explicit Euler method as our running example.

\begin{figure}[!t]
\centering
\subfigure[Euler]{\includegraphics[width=0.41\columnwidth]{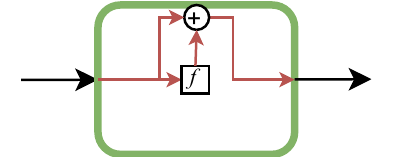}}
\subfigure[RK4]{\includegraphics[width=0.57\columnwidth]{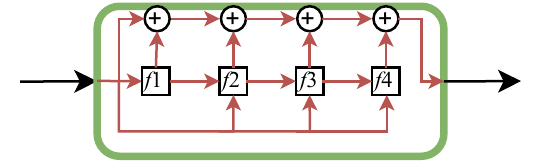}}
\caption{The explicit Euler method and RK4 in a step. To derive $\bm{h}(t + s)$ from $\bm{h}(t)$ with a step size $s$, RK4 is four times more complicated than the explicit Euler method. Note that the explicit Euler method is the same as the residual connection and RK4 is the same as the dense connection when $f$ is a neural network layer.}\label{fig:solvers}
\end{figure}

One more ODE solver that is worth mentioning is the implicit Adams–Moulto method which is written as follows:
\begin{linenomath*}\begin{align}\label{eq:adam}
\bm{h}(t + s) = \bm{h}(t) + \frac{s}{24}\Big(9f_1 + 19f_2 - 5f_3 + f_4\Big),
\end{align}\end{linenomath*}where $f_1 = f(\bm{h}(t+s), t+s;\bm{\theta}_f)$, $f_2 = f(\bm{h}(t), t;\bm{\theta}_f)$, $f_3 = f(\bm{h}(t-s), t-s;\bm{\theta}_f)$, and $f_4 = f(\bm{h}(t-2s), t-2s;\bm{\theta}_f)$. Both $f_3$ and $f_4$ are from previous history and we do not need to newly evaluate them.

This implicit method is different from the aforementioned explicit solvers in that i) it uses past multi-step history, i.e., $f_3$ and $f_4$, to calculate a more robust derivative term and ii) it uses $f_1$ in conjunction with the multi-step history. At the moment of time $t$, however, it is before calculating $\bm{h}(t+s)$ so evaluating $f_1$ cannot be done in a naive way. It uses advanced methods, such as Newton's method, to solve for $\bm{h}(t + s)$. The use of $f_1$ is called \emph{implicit} in the field of numerical methods to solve ODEs. Its analogy to neural network connection has not been studied yet due to the implicit nature of the method. However, it falls into the category of dense networks because it uses multi-step information.

For our experiments, we consider all those advanced solvers, which is one more advantage of our formulating the graph-based CF as the dual co-evolving ODEs.


\subsection{Collaborative Filtering (CF)} Let $\bm{E}^u_0 \in \mathbb{R}^{N \times D}$ and $\bm{E}^p_0 \in \mathbb{R}^{M \times D}$ be the initial user and product embeddings, respectively. There are $N$ users and $M$ products, and embeddings are $D$ dimensions. Early CF methods include matrix factorization~\cite{5197422}, SVD++~\cite{10.1145/1401890.1401944}, neural attentive item similarity~\cite{8352808}, and so on. All these methods utilize user-product interaction history~\cite{10.1145/3285029}.


Because user-product relationships can be represented by bipartite graphs, it recently became popular to adopt GCNs for CF~\cite{10.1145/3331184.3331267,Chen_Wu_Hong_Zhang_Wang_2020,10.1145/3397271.3401063}. NGCF is one of the most popular GCN-based CF methods. It uses non-linear activations and transformation matrices to transform from the user embedding space to the product embedding space, and vice versa. At each layer, user and product embeddings are extracted as in the layer combination. However, it concatenates them instead of taking their sum. Its overall architecture is similar to the standard GCN~\cite{kipf2017semi}. However, it was later noted that the adoption of the non-linear activation and the embedding space transformation are not necessary in CF due to the environmental dissimilarity between general graph-based downstream tasks and CF~\cite{10.1145/3397271.3401063}. That is, other graph-based tasks include abundant information, e.g., high-dimensional node features. However, CF frequently includes a bipartite graph without additional features. Even worse, the graph is sparse in CF. Due to the difference, non-linear GCNs are easily overfitted to training graphs and their testing accuracy is mediocre in many cases even with various countermeasures preventing it. It was also empirically proven that transforming between user and product embedding spaces is not helpful in CF~\cite{10.1145/3397271.3401063}.

After NGCF, several methods have been proposed. Among them, in particular, one recent graph-based method, called LightGCN, shows state-of-the-art accuracy in many datasets. In addition, it also showed that linear GCNs with layer combination work the best among many design choices. Its linear graph convolutional layer definition is as follows:
\begin{linenomath*}\begin{align}\begin{split}\label{eq:lgc}
    \bm{E}^u_k = \tilde{\bm{A}}_{u\rightarrow p}\bm{E}^p_{k-1},\\
    \bm{E}^p_k = \tilde{\bm{A}}_{p\rightarrow u}\bm{E}^u_{k-1},
\end{split}\end{align}\end{linenomath*}where $\tilde{\bm{A}}_{p\rightarrow u} \in [0,1]^{M \times N}$ is a normalized adjacency matrix of the graph from products to users and $\tilde{\bm{A}}_{u\rightarrow p} \in [0,1]^{N \times M}$ is also defined in the same way but from users to products. LightGCN learns the initial embeddings, denoted $\bm{E}^u_0$ and $\bm{E}^p_0$, and uses the layer combination, which can be written as follows:
\begin{linenomath*}\begin{align}\begin{split}\label{eq:comb}
    \bm{E}^u_{final} = \sum_{i=0}^{K}w_i\bm{E}^u_i,\\
    \bm{E}^p_{final} = \sum_{i=0}^{K}w_i\bm{E}^p_i,
\end{split}\end{align}\end{linenomath*}where $K$ is the number of layers, $w_i$ is a coefficient, and $\bm{E}^u_{final}$ and $\bm{E}^p_{final}$ are the final embeddings.

The CF methods, including our method, LightGCN, and so on, learn the initial embeddings of users and products (and model parameters if any). After a series of $K$ graph convolutional layers, a graph-based CF algorithm derives $\bm{E}^u_{final}$ and $\bm{E}^p_{final}$ and use their dot products to predict $r_{u,i}$, a rating (or ranking score) by user $u$ to product $i$, for all $u,i$. Ones typically use the following Bayesian personalized ranking (BPR) loss~\cite{10.5555/1795114.1795167} to train the initial embedding vectors (and model parameters if any) in the field of CF:
\begin{linenomath*}\begin{align}\label{eq:bpr}
L = -\sum_{u=1}^{N}\sum_{i \in \mathcal{N}_u}\sum_{j \notin \mathcal{N}_u} \ln \sigma (r_{u,i} - r_{u,j}) + \lambda \| \bm{E}^u_0 \odot \bm{E}^p_0 \|^2,
\end{align}\end{linenomath*}where $\mathcal{N}_u$ is a set of products neighboring to $u$, $\sigma$ is a non-linear activation, and $\odot$ means the concatenation operator. We use the softplus for $\sigma$.

\subsection{Linear GCNs and Newton's Law of Cooling}\label{sec:thermal}
As a matter of fact, Eq.~\eqref{eq:lgc} is similar to the heat equation, which describes the law of thermal diffusive processes, i.e., Newton's Law of Cooling. The heat equation can be written as follows:
\begin{align}\begin{split}\label{eq:heat}
    \frac{d \bm{H}_{t}}{dt} = -\Delta \bm{H}_{t},
\end{split}\end{align}where $\Delta$ is the Laplace operator and $\bm{H}_{t}$ is a column vector which contains the temperatures of the nodes in a graph or a discrete grid at time $t$. The Laplace operator $\Delta$ is simply a matrix multiplication with the Laplacian matrix or the normalized adjacency matrix.

Therefore, the right-hand side of Eq.~\eqref{eq:lgc} can be reduced to Eq.~\eqref{eq:heat} if we interpret each element of $\bm{E}^u_i$ and $\bm{E}^p_i$ as a temperature value --- since they are $D$-dimensional vectors, we can consider that $D$ different diffusive processes exist in Eq.~\eqref{eq:lgc}. In this regard, we can consider that LightGCN models discrete thermal diffusive processes whereas our method describes continuous thermal diffusive processes.

\section{Proposed Method}


In this section, we describe our proposed method. Our main idea is to design co-evolutionary ODEs of user and product embeddings with a continuous and learnable time variable $t$. 


\subsection{Overall Architecture}
In Fig.~\ref{fig:archi} (b), we show the overall architecture of LT-OCF. The two initial embeddings, $\bm{E}^u_0$ and $\bm{E}^p_0$, are fed into the dual co-evolutionary ODEs. Then, we have the layer combination architecture to derive the final embeddings. The distinguished feature of LT-OCF lies in the dual ODE layer, where we can interpret the time variable $t$ as a continuous layer variable.

LT-OCF enjoys the continuous characteristic of $t$ and construct a more flexible architecture. In LightGCN and other existing GCN-based CF methods, we have to use pre-determined discrete architectures. However, LT-OCF can use any positive real numbers for $t$ and those numbers are even trainable in our case.

\subsection{ODE-based User and Product Embeddings} The user and product embedding co-evolutionary processes can be written as follows:
\begin{linenomath*}\begin{align}\begin{split}\label{eq:panode}
    \bm{u}(K) =&\; \bm{u}(0) + \int_{0}^{K}f(\bm{p}(t))dt,\\
    \bm{p}(K) =&\; \bm{p}(0) + \int_{0}^{K}g(\bm{u}(t))dt,
\end{split}\end{align}\end{linenomath*}where $\bm{u}(t) \in \mathbb{R}^{N \times D}$ is a user embedding matrix and $\bm{p}(t) \in \mathbb{R}^{M \times D}$ is a product embedding matrix at time $t$. $f(\bm{p}(t))$ outputs $\frac{d \bm{u}(t)}{dt}$ and $g(\bm{u}(t))$ outputs $\frac{d \bm{p}(t)}{dt}$. $\bm{u}(0) = \bm{E}^u_0$ and $\bm{p}(0) = \bm{E}^p_0$ in our case because the initial embeddings are directly fed into the ODEs (cf. Fig.~\ref{fig:archi} (b)). We note that $\bm{u}(t)$ and $\bm{p}(t)$ constitute a set of co-evolving ODEs. User embedding influences product embedding and vice versa. Therefore, our co-evolving ODEs are a reasonable design choice.

However, this formulation does not fully describe our proposed concept of \emph{learnable-time} and we propose a more advanced formulation in the next paragraph. 

\subsubsection{Learnable-time Architecture.} 
\begin{figure}[]
\centering
\includegraphics[width=1\columnwidth]{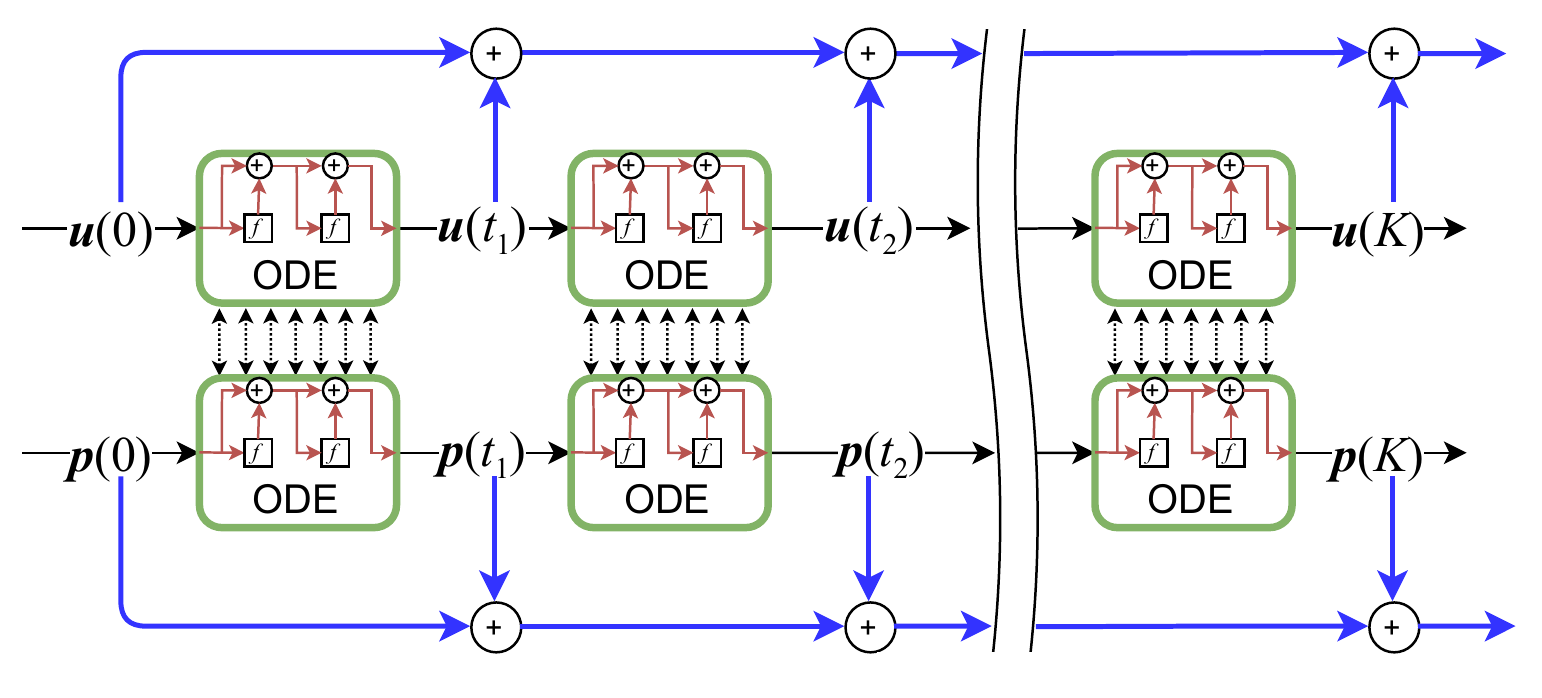}
\caption{Our proposed dual co-evolving ODEs. In this figure, we assume the explicit Euler method. The residual connections inside the ODEs are in red and the layer combination is in blue. The time-points $\{t_1, \cdots, t_T\}$ to construct the layer combination are trained in our framework. In this figure, each ODE internally uses two steps (i.e., two red residual connections) to solve an integral problem but this is only for illustration purposes. In real environments, the number of steps for each ODE can be varied.} \label{fig:dual}
\end{figure}

In our framework, we can learn how to construct the layer combination (rather than relying on a manually designed architecture). In order to adopt such an advanced option, we extract $\bm{u}(t)$ and $\bm{p}(t)$ with several different learnable time-points $t \in \{t_1, \cdots, t_T\}$, where $T$ is a hyperparameter, and $0 < t_i < t_{i+1} < K$ for all $i$. Therefore, Eq.~\eqref{eq:panode} can be re-written as follows:
\begin{linenomath*}\begin{align}\begin{split}\label{eq:panode2}
    \bm{u}(t_1) =&\; {\color{red}\bm{u}(0) +} \int_{0}^{t_1}f(\bm{p}(t))dt,\\
    \bm{p}(t_1) =&\; {\color{red}\bm{p}(0) +} \int_{0}^{t_1}g(\bm{u}(t))dt,\\
    \vdots\\
    \bm{u}(K) =&\; {\color{red}\bm{u}(t_T) +} \int_{t_T}^{K}f(\bm{p}(t))dt,\\
    \bm{p}(K) =&\; {\color{red}\bm{p}(t_T) +} \int_{t_T}^{K}g(\bm{u}(t))dt,
\end{split}\end{align}\end{linenomath*}where $t_i$ is trainable for all $i$. The parts of the equation highlighted in red are used to create residual connections (cf. the red residual connections inside the ODEs in Fig.~\ref{fig:dual}). The final embeddings are calculated as follows:
\begin{linenomath*}\begin{align}\begin{split}\label{eq:comb2}
    \bm{E}^u_{final} = w_0\bm{u}(0) + \sum_{i=1}^{T}w_i\bm{u}(t_i) + w_K\bm{u}(K),\\
    \bm{E}^p_{final} = w_0\bm{p}(0) + \sum_{i=1}^{T}w_i\bm{p}(t_i) + w_K\bm{p}(K).
\end{split}\end{align}\end{linenomath*}

Recall that what the explicit Euler method does internally is to generalize residual connections in a continuous manner. So, extracting intermediate ODE states (i.e., $\bm{u}(t)$ and $\bm{p}(t)$ with $t \in \{t_1, \cdots, t_T\}$) and creating a higher level of layer combination in Eq.~\eqref{eq:comb2} correspond to dual residual connections (cf. the blue layer combination outside the ODEs in Fig.~\ref{fig:dual}).


If using other advanced ODE solvers, the connections inside the ODEs become more sophisticated, e.g., DenseNet or FractalNet connections if RK4 is used. In Table~\ref{tbl:ode}, we summarize all those cases.

\begin{table}[t]
\setlength{\tabcolsep}{3pt}
\centering
\small
\caption{Various network architectures (inside and outside the ODEs in Fig.~\ref{fig:dual}) created by various ODE solvers}\label{tbl:ode}
\begin{tabular}{c|c|c}
\specialrule{1pt}{1pt}{1pt}
ODE Solver & Inside the ODEs & Outside the ODEs \\ \specialrule{1pt}{1pt}{1pt}
\begin{tabular}[c]{@{}c@{}}Explicit Euler\\(without the red parts in Eq.~\eqref{eq:panode2})\end{tabular}  & \begin{tabular}[c]{@{}c@{}}Linear\\Connection\end{tabular}  & \multirow{9}{*}{\begin{tabular}[c]{@{}c@{}}Layer\\Combination\end{tabular}} \\  \cline{1-2}
\begin{tabular}[c]{@{}c@{}}Explicit Euler\\(with the red parts in Eq.~\eqref{eq:panode2})\end{tabular} & \begin{tabular}[c]{@{}c@{}}Residual\\Connection\end{tabular}  &  \\ \cline{1-2}
RK4 & \begin{tabular}[c]{@{}c@{}}DenseNet or\\FractalNet\end{tabular} &  \\\cline{1-2}
Adams-Moulto & \begin{tabular}[c]{@{}c@{}}Implicit\\Connection\end{tabular} &  \\ \cline{1-2}
DOPRI & \begin{tabular}[c]{@{}c@{}}Adaptive\\Connection\end{tabular}  &  \\ \specialrule{1pt}{1pt}{1pt}
\end{tabular}
\end{table}

\subsubsection{Non-parameterized and Non-time-dependent ODEs.} We need to define the two ODE functions, $f$ and $g$. Being inspired by the recent success of linear graph convolutions, we use the following definition for $f$ and $g$:
\begin{linenomath*} \begin{align}\begin{split}\label{eq:odef}
f(\bm{p}(t)) &= \bar{\bm{A}}_{u\rightarrow p}\bm{p}(t),\\
g(\bm{u}(t)) &= \bar{\bm{A}}_{p\rightarrow u}\bm{u}(t),
\end{split}\end{align}\end{linenomath*}where $\bar{\bm{A}}$ means either the symmetric normalized Laplacian matrix or the normalized adjacency matrix. LightGCN uses the latter but our method based on the continuous thermal diffusive differential equation can use both of them.

We note that our definitions for $f$ and $g$ will result in non-parameterized and non-time-dependent ODEs because our ODE functions do not require $t$, $\bm{\theta}_f$, and $\bm{\theta}_g$ as their input.



\subsubsection{Relation with Linear GCN-based CF Methods} There exist several linear GCNs. LightGCN studied about the similarity among various such linear GCN models and showed many other linear models can be approximated as a special case of LightGCN. In this subsection, we study about the similarity between our method and LightGCN.

Suppose the following setting in our method: i) $t_i$ is not trained but fixed to $i$ for all $i$, ii) We use the explicit Euler method with its step size parameter $s = 1$, and iii) We do not use the residual connection but the linear connection inside the ODEs, i.e., removing the red parts in Eq.~\eqref{eq:panode2}. This specific setting can be written as follows:
\begin{linenomath*}\begin{align}\begin{split}\label{eq:panode3}
    \bm{u}(1) =&\; f(\bm{p}(0)),\\
    \bm{p}(1) =&\; g(\bm{u}(0)),\\
    \vdots\\
    \bm{u}(K) =&\; f(\bm{p}(K-1)),\\
    \bm{p}(K) =&\; g(\bm{u}(K-1)).
\end{split}\end{align}\end{linenomath*}

After that, the linear combination yields $\bm{E}^u_{final} = \sum_{i=0}^{K}w_i\bm{u}(t_i)$ and $\bm{E}^p_{final} = \sum_{i=0}^{K}w_i\bm{p}(t_i)$. We note that these final embeddings are equivalent to Eq.~\eqref{eq:comb} because our ODE functions $f$ and $g$ in Eq.~\eqref{eq:odef} are equivalent to the linear layer definition of LightGCN in Eq.~\eqref{eq:lgc}. Thus, our method is equivalent to LightGCN under the specific setting. Therefore, one can consider our method, LT-OCF, as a \emph{continuous} generalization of linear GCNs, including LightGCN and others that can be approximated by LightGCN.

\subsection{How to Train.}\label{sec:tra}
Our proposed method includes a couple of sophisticated techniques and its training algorithm is inevitably more complicated than other cases. We also use the BPR loss, denoted $L$, to train our model, which is common for many CF methods.

We propose to alternately train the co-evolving ODEs and their intermediate time points. When training for one, we fix all other parts. This makes the gradient calculation by the adjoint sensitivity method simple because a fixed ODE/time point can be considered as constants at a moment of training time. The gradients of loss w.r.t. $\bm{u}(0)$, which is identical to the initial embedding $\bm{E}^u_0$, can be calculated via the following reverse-mode integration~\cite{NIPS2018_7892}:
\begin{linenomath*}\begin{align}\begin{split}
    \frac{d L}{d \bm{u}(0)} =  \bm{a}_{\bm{u}}(K) - \int_{K}^{0}\bm{a}_{\bm{u}}(t)^\intercal\frac{\partial g(\bm{u}(t))}{\partial \bm{u}(t)},
\end{split}\end{align}\end{linenomath*}where $\bm{a}_{\bm{u}}(t) \stackrel{\text{def}}{=} \frac{d L}{d \bm{u}(t)}$. The gradients of loss w.r.t. $\bm{p}(0)$, the initial embedding of products, can be done in the same way and we omit its description for space reasons. Calculating the gradients requires a space complexity of $\mathcal{O}(1)$ and a time complexity of $\mathcal{O}(\frac{1}{s})$, where $s$ is the average step-size of underlying ODE solver which is fixed for the Euler method and RK4 and varied for DOPRI, because we use the adjoint sensitivity method.

The gradient of loss w.r.t. $t_i$ does not involve the adjoint sensitivity method but is defined directly as follows:
\begin{linenomath*}\begin{align}\begin{split}
\frac{d L}{d t_i} =& \frac{\partial L}{\partial \bm{u}(t_i)}\frac{d \bm{u}(t_i)}{d t_i} + \frac{\partial L}{\partial \bm{p}(t_i)}\frac{d \bm{p}(t_i)}{d t_i}\\
=& \bm{a}_{\bm{u}}(t_i)g(\bm{u}(t_i)) + \bm{a}_{\bm{p}}(t_i)f(\bm{p}(t_i)),
\end{split}\end{align}\end{linenomath*}where its complexity is $\mathcal{O}(T)$ to train all time-points.

\begin{algorithm}[t]
\SetAlgoLined
\caption{How to train $\bm{E}^u_{0}$ and $\bm{E}^p_{0}$}\label{alg:train}
\KwIn{Rating matrix $\bm{R}$}
Initialize $\bm{E}^u_{0}$ and $\bm{E}^p_{0}$;

\While {the BPR loss $L$ is not converged}{
    Update $\bm{E}^u_{0}$ with $\frac{d L}{d \bm{u}(0)}$;
    
    Update $\bm{E}^p_{0}$ with $\frac{d L}{d \bm{p}(0)}$;
    
    
    Update $t_i$ for all $i$\;
    
}
\Return $\bm{E}^u_{0}$ and $\bm{E}^p_{0}$;
\end{algorithm}

Our propose training algorithm is in Alg.~\ref{alg:train}. We alternately train each part until the BPR loss converges.

\subsubsection{On the Tractability of Training.}\label{sec:well}
The Cauchy--Kowalevski theorem states that, given $f = \frac{d \bm{h}(t)}{d t}$, there exists a unique solution of $\bm{h}$ if $f$ is analytic (or locally Lipschitz continuous), i.e. the ODE problem is well-posed if $f$ is analytic~\cite{10.2307/j.ctvzsmfgn}. In our case, Eq.~\eqref{eq:odef}, which is to model $\frac{d \bm{u}(t)}{d t}$ and $\frac{d \bm{p}(t)}{d t}$, uses matrix multiplications that are analytic. This implies that there will be only a unique optimal ODE for $\bm{u}(t)$, given fixed $\bm{p}(t)$ and vice versa. Because of i) the uniqueness of the solution and ii) our relatively simpler definitions of $f$ and $g$ in comparison with other NODE applications, we believe that our training method can find a good solution.

\section{Experimental Evaluations}
In this section, we introduce our experimental environments and results. All experiments were conducted in the following software and hardware environments: \textsc{Ubuntu} 18.04 LTS, \textsc{Python} 3.6.6, \textsc{Numpy} 1.18.5, \textsc{Scipy} 1.5, \textsc{Matplotlib} 3.3.1, \textsc{PyTorch} 1.2.0, \textsc{CUDA} 10.0, and \textsc{NVIDIA} Driver 417.22, i9 CPU, and \textsc{NVIDIA RTX Titan}. Our source codes and data are at {\color{blue}\url{https://github.com/jeongwhanchoi/LT-OCF}}.

\begin{table}[t]
\centering
\setlength{\tabcolsep}{4pt}
\caption{Statistics of datasets}\label{tbl:data}
\begin{tabular}{cccc}
\specialrule{1pt}{1pt}{1pt}
Name & \#Users & \#Items & \#Interactions\\ \specialrule{1pt}{1pt}{1pt}
Gowalla & 29,858 & 40,981 & 1,027,370 \\
Yelp2018 & 31,668 & 38,048 & 1,561,406 \\
Amazon-Book & 52,643 & 91,599 & 2,984,108 \\
\specialrule{1pt}{1pt}{1pt}
\end{tabular}
\end{table}

\subsection{Experimental Environments}
\subsubsection{Datasets and Baselines.} We use the three benchmark datasets used by previous works without any modifications: Gowalla, Yelp2018, and Amazon-Book~\cite{10.1145/3331184.3331267,Chen_Wu_Hong_Zhang_Wang_2020,10.1145/3397271.3401063}. Their statistics are summarized in Table~\ref{tbl:data}. We consider the following baselines to compare with:
\begin{enumerate}
    \item MF~\cite{10.5555/1795114.1795167} is a matrix decomposition optimized by Bayesian Personalization Rank (BPR) loss, which utilizes the user-item direct interaction only as the target value of the interaction function.
    \item Neu-MF is a neural collaborative filtering method~\cite{He2017}. This method uses multiple hidden layers above the element-wise concatenation of user and item embeddings to capture their non-linear feature interactions.
    \item CMN~\cite{Ebesu2018} is a state-of-the-art memory-based model. This method uses first-order connections to find similar users who interacted with the same items.
    \item HOP-Rec~\cite{yang2018hop} is a graph-based model, which uses the high-order user-item interactions by random walks to enrich the original training data.
    \item GC-MC~\cite{DBLP:journals/corr/BergKW17} is a graph auto-encoder framework based on differentiable message passing on the bipartite interaction graph. This method applies the GCN encoder on user-item bipartite graph and employs one convolutional layer to exploit the direct connections between users and items.
    \item Mult-VAE is a variational autoencoder-based CF method~\cite{10.1145/3178876.3186150}. We use a drop-out rate of \{0, 0.2, 0.5\} and $\beta$ of \{0.2, 0.4, 0.6, 0.8\}. The layer-wise dimensionality is 600, 200, and then 600 as recommended in the paper and the authors.
    \item GRMF is a matrix factorization method by adding the graph Laplacian regularizer~\cite{NIPS2015_f4573fc7}. We change the original loss of GRMF to the BPR loss for fair comparison. GRMG-norm is a slight variation from GRMF by adding a normalization to graph Laplacian.
    \item NGCF~\cite{10.1145/3331184.3331267} is a representative GCN-based CF method. It uses feature transformation and non-linear activations.
    \item LR-GCCF~\cite{Chen_Wu_Hong_Zhang_Wang_2020} and LightGCN~\cite{10.1145/3397271.3401063} are linear GCN-based CF methods. They currently show state-of-the-art accuracy.
\end{enumerate}

We use the two standard evaluation metrics, Recall@20 and NDCG@20, with the all-ranking protocol, i.e., all items that do not have any interactions with a user are recommendation candidates.

\subsubsection{Hyperparameters.} Our method and the above baseline models have several common hyperparameters. In this paragraph, we introduce them.
\begin{enumerate}
\item The regularization coefficient $\lambda$ in all methods is in \{\num{1.0e-4}, \num{1.0e-3}, \num{1.0e-2}\}.
\item The dimensionality of embedding vectors is 64 as recommended in~\cite{10.1145/3397271.3401063}, and a Normal distribution of $\mathcal{N}(\bm{0}, \bm{0.1})$ is used to set initial embeddings.
\item The layer combination coefficient $w_i = \frac{1}{1+K}$, where $K$ is the number of elements in the layer combination.
\item The number of elements $K$ is in \{2,3,4\}.
\item The number of learnable intermediate time points $T$ is in \{1,2,3\}.
\item We use the same early stopping criterion as that of NGCF and train with Adam and a learning rate in \{\num{1.0e-5}, \num{1.0e-4}, \num{1.0e-3}, \num{1.0e-2}\}.
\item We consider the following ODE solvers: the explicit Euler method, RK4, Adams-Moulto, and DOPRI.
\end{enumerate}

The best configuration set in each data is as follows: In Gowalla, $\lambda=\num{1.0e-4}$, learning rate $=\num{1.0e-4}$, learning rate for time $=\num{1.0e-6}$, $K=4, T=3$; In Yelp2018, $\lambda=\num{1.0e-4}$, learning rate $=\num{1.0e-5}$, learning rate for time $=\num{1.0e-6}$,  $K=4, T=3$; In Amazon-Book, $\lambda=\num{1.0e-4}$, learning rate $=\num{1.0e-4}$, learning rate $=\num{1.0e-6}$, $K=4, T=3$.



\subsection{Experimental Results}

\begin{table}[t]
\centering
\small
\setlength{\tabcolsep}{3pt}
\caption{Model performance comparison. LT-OCF significantly outperforms other methods in Amazon-Book.}\label{tbl:acc}
\begin{tabular}{ccccccc}
\specialrule{1pt}{1pt}{1pt}
Dataset & \multicolumn{2}{c}{Gowalla} & \multicolumn{2}{c}{Yelp2018} & \multicolumn{2}{c}{Amazon-Book} \\
Method & Recall & NDCG & Recall & NDCG & Recall & NDCG \\ \specialrule{1pt}{1pt}{1pt}
MF & 0.1291 & 0.1109 & 0.0433 & 0.0354 & 0.0250 & 0.0196 \\
NeuMF & 0.1399 & 0.1212 & 0.0451 & 0.0363 & 0.0258 & 0.0200 \\
CMN & 0.1405 & 0.1221 & 0.0475 & 0.0369 & 0.0267 & 0.0218 \\
HOP-Rec & 0.1399 & 0.1214 & 0.0517 & 0.0428 & 0.0309 & 0.0232 \\
GC-MC & 0.1395 & 0.1204 & 0.0462 & 0.0379 & 0.0288 & 0.0224 \\
PinSage & 0.1380 & 0.1196 & 0.0471 & 0.0393 & 0.0282 & 0.0219 \\
Mult-VAE & 0.1641 & 0.1335 & 0.0584 & 0.0450 & 0.0407 & 0.0315 \\
GRMF & 0.1477 & 0.1205 & 0.0571 & 0.0462 & 0.0354 & 0.0270 \\
GRMF-Norm & 0.1557 & 0.1261 & 0.0561 & 0.0454 & 0.0352 & 0.0269 \\
NGCF & 0.1570 & 0.1327 & 0.0579 & 0.0477 & 0.0344 & 0.0263 \\
LR-GCCF & 0.1518 & 0.1259 & 0.0574 & 0.0349 & 0.0341 & 0.0258 \\
LightGCN & 0.1830 & 0.1554 & 0.0649 & 0.0530 & 0.0411 & 0.0315 \\\hline
LT-OCF & \textbf{0.1875} & \textbf{0.1574} & \textbf{0.0671} & \textbf{0.0549} & \textbf{0.0442} & \textbf{0.0341} \\ \specialrule{1pt}{1pt}{1pt}
\end{tabular}
\end{table}

In Table~\ref{tbl:acc}, we summarize the overall accuracy in terms of recall and NDCG. The non-linear GCN-based method, NGCF, shows good performance for a couple of cases in comparison with other non-GCN-based methods. After that, LightGCN shows the state-of-the-art accuracy in all cases among all baselines. It sometimes outperforms other methods by large margins, e.g., a recall of 0.1830 in Gowalla by Light GCN vs. a recall of 0.1641 by Multi-VAE. In general, the three GCN-base methods, NGCF, LR-GCCF, and LightGCN, outperform other baseline methods by large margins.


However, the best accuracy is straightly marked by our method, LT-OCF, in all cases. All those best results are achieved by RK4, which implies that the linear GCN architecture of LightGCN may not be the best option (see our discussion in Section~\ref{sec:dis}). In particular, our method's NDCG in Amazon-Book shows an improvement of approximately 10\% over LightGCN.

\begin{figure}[t]
\centering
\subfigure[Training curve of loss]{\includegraphics[width=0.49\columnwidth]{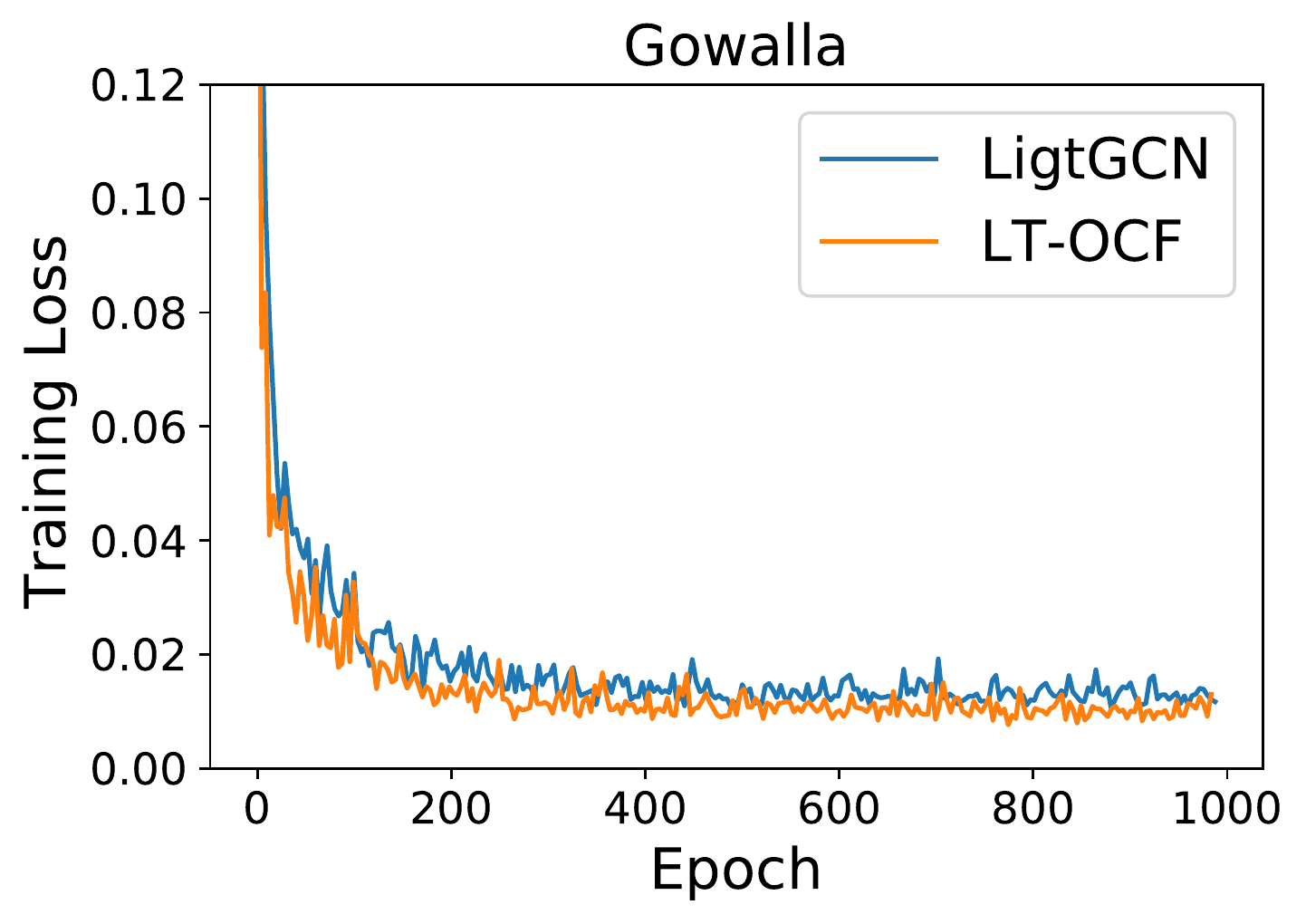}}
\subfigure[Training curve of recall]{\includegraphics[width=0.49\columnwidth]{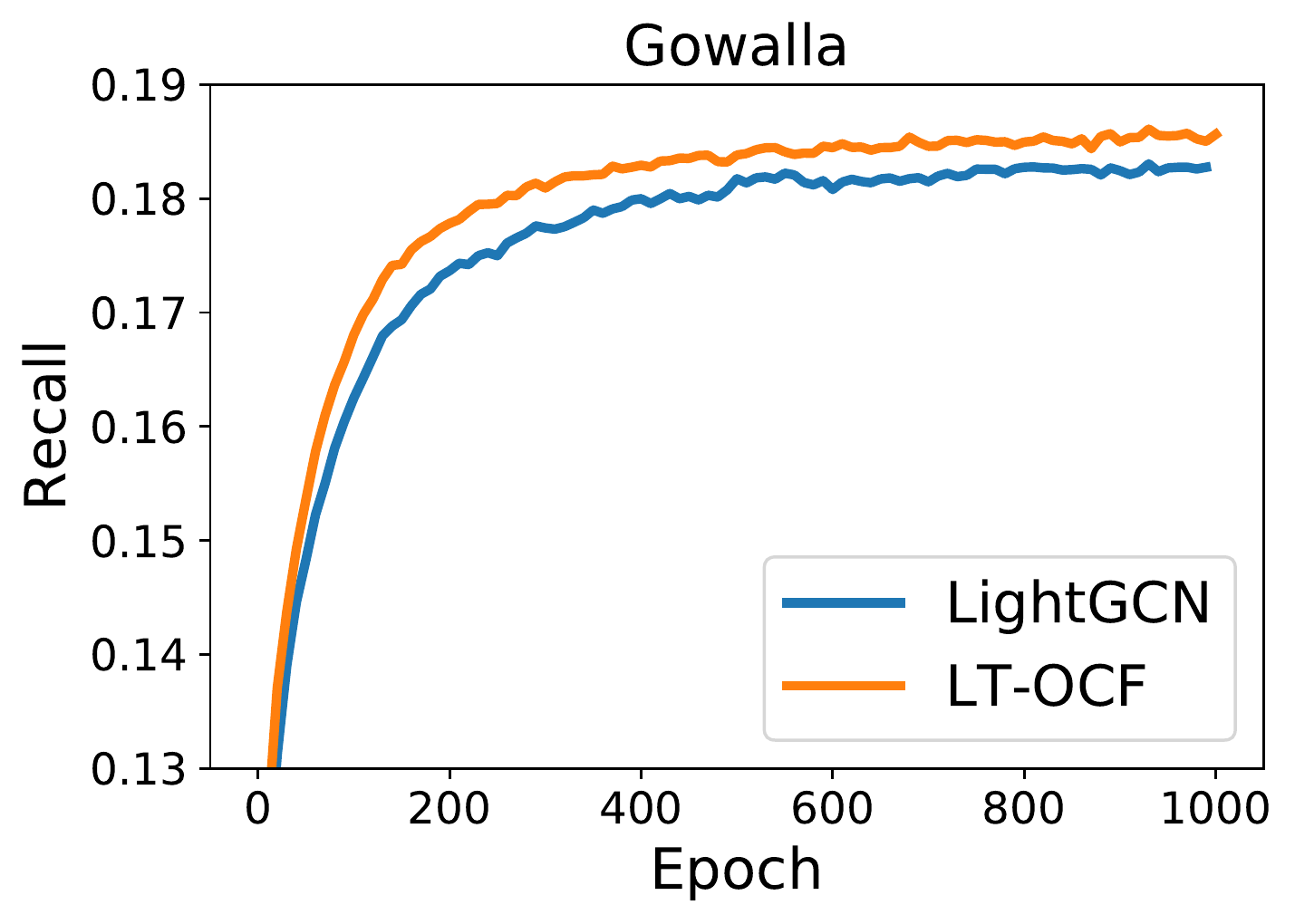}}
\subfigure[Training curve of NDCG]{\includegraphics[width=0.49\columnwidth]{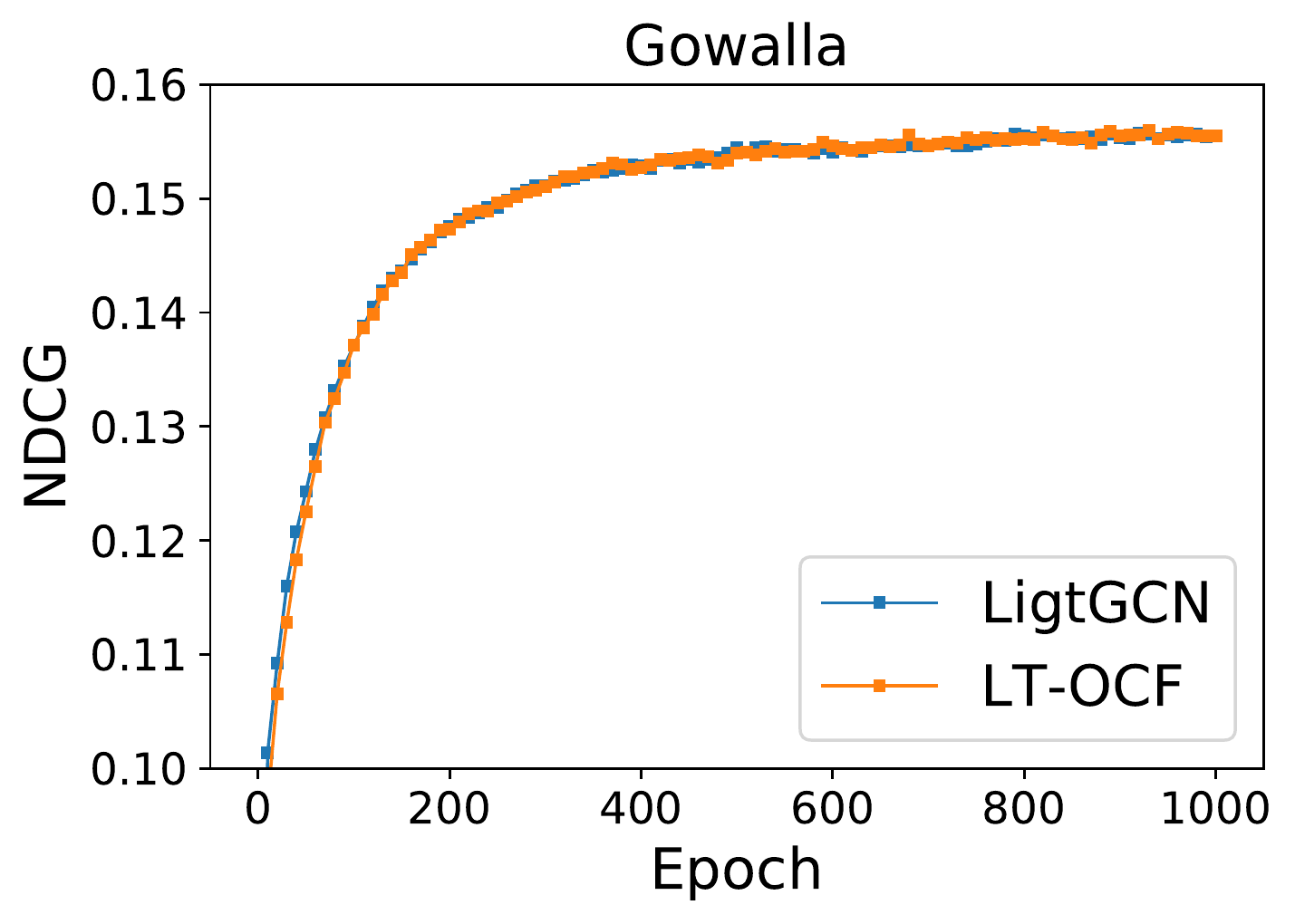}}
\subfigure[Training curve of $t_i$]{\includegraphics[width=0.49\columnwidth]{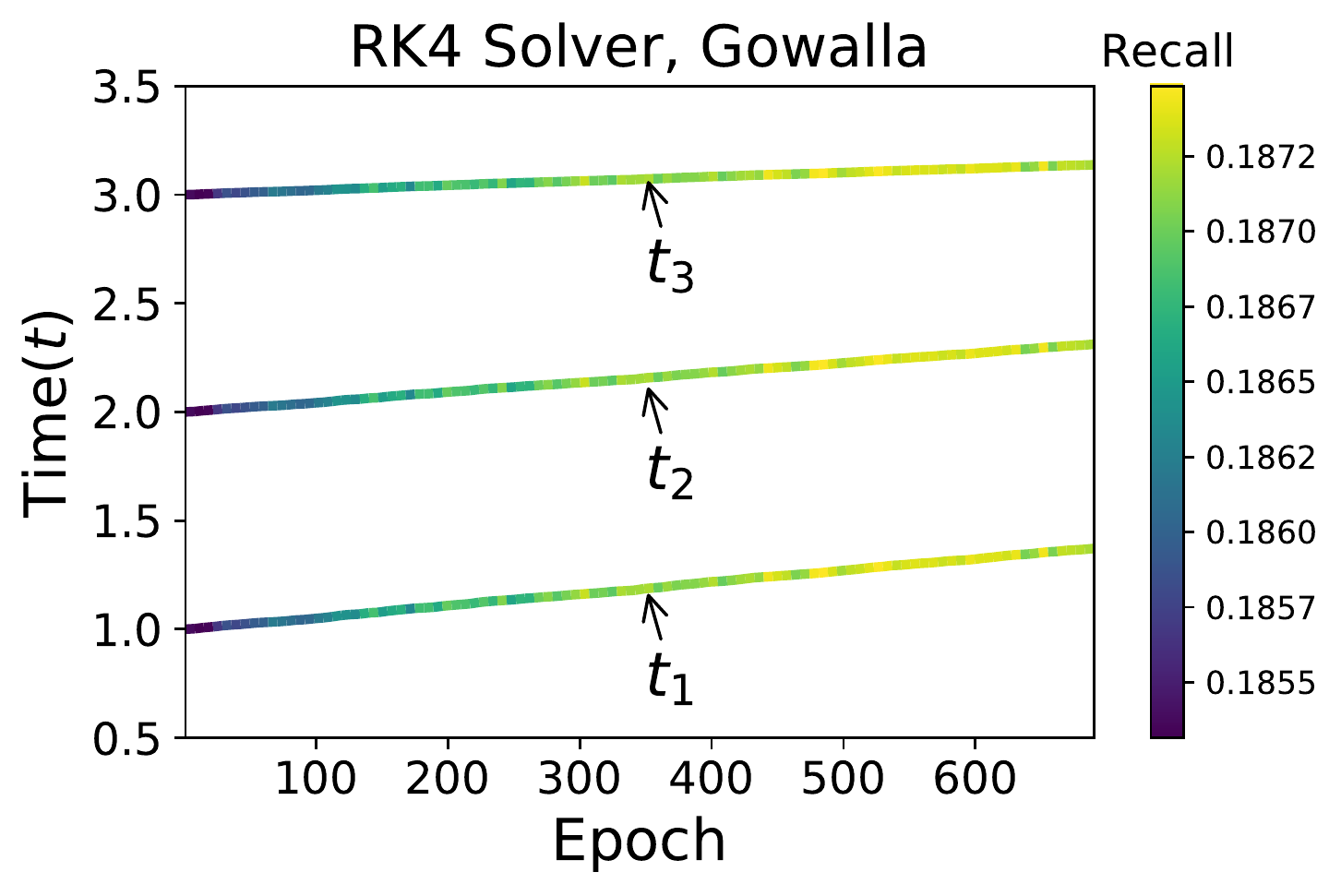}}

\caption{Training curves in Gowalla. (a) The loss curves, (b) The training curves of Recall, (c) The training curves of NDCG, (d) The training curves of $t_1, t_2, t_3$ when $T=3$.} \label{fig:time1}
\end{figure}

\begin{figure}[t]
\centering
\subfigure[Training curve of loss]{\includegraphics[width=0.49\columnwidth]{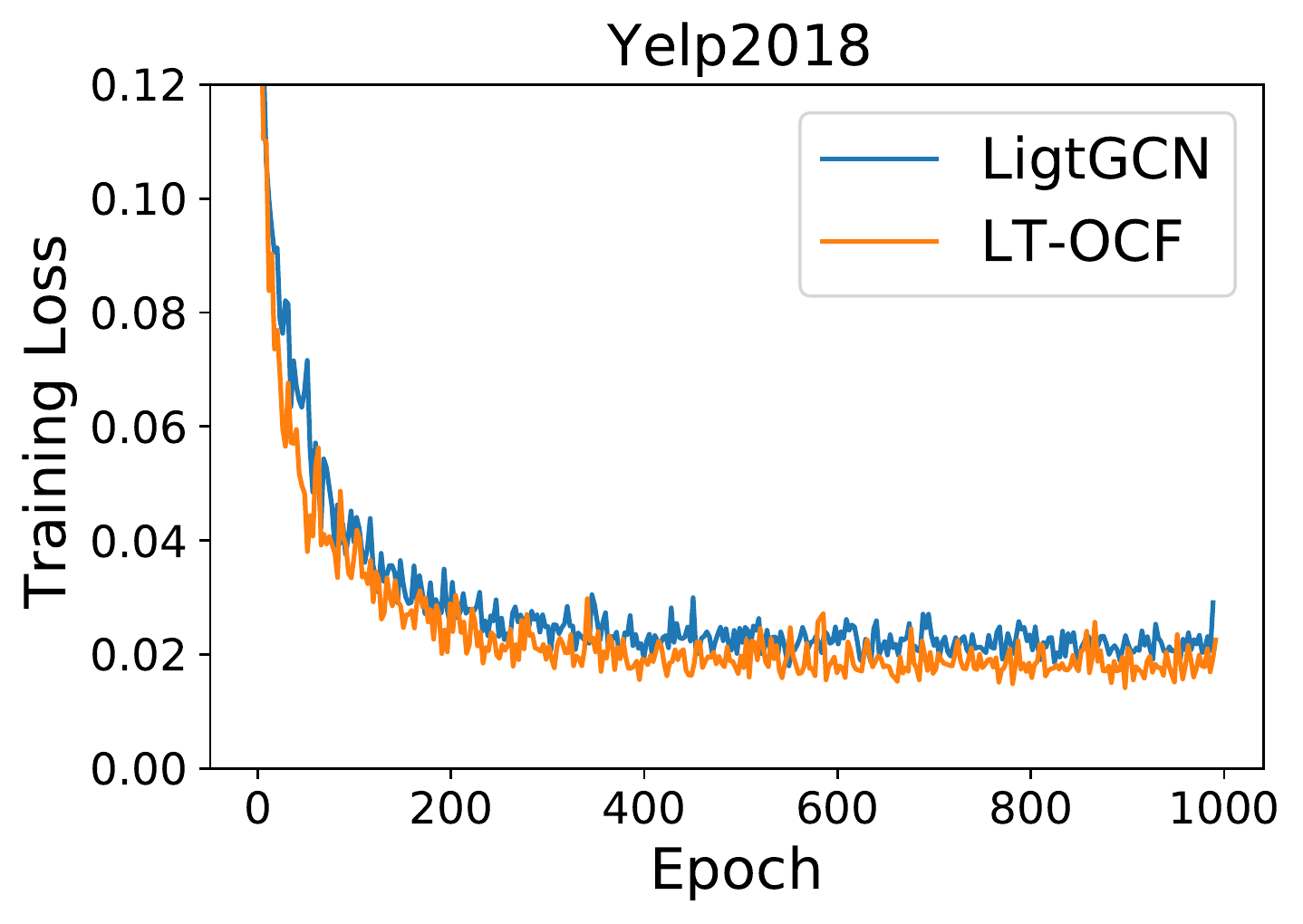}}
\subfigure[Training curve of recall]{\includegraphics[width=0.49\columnwidth]{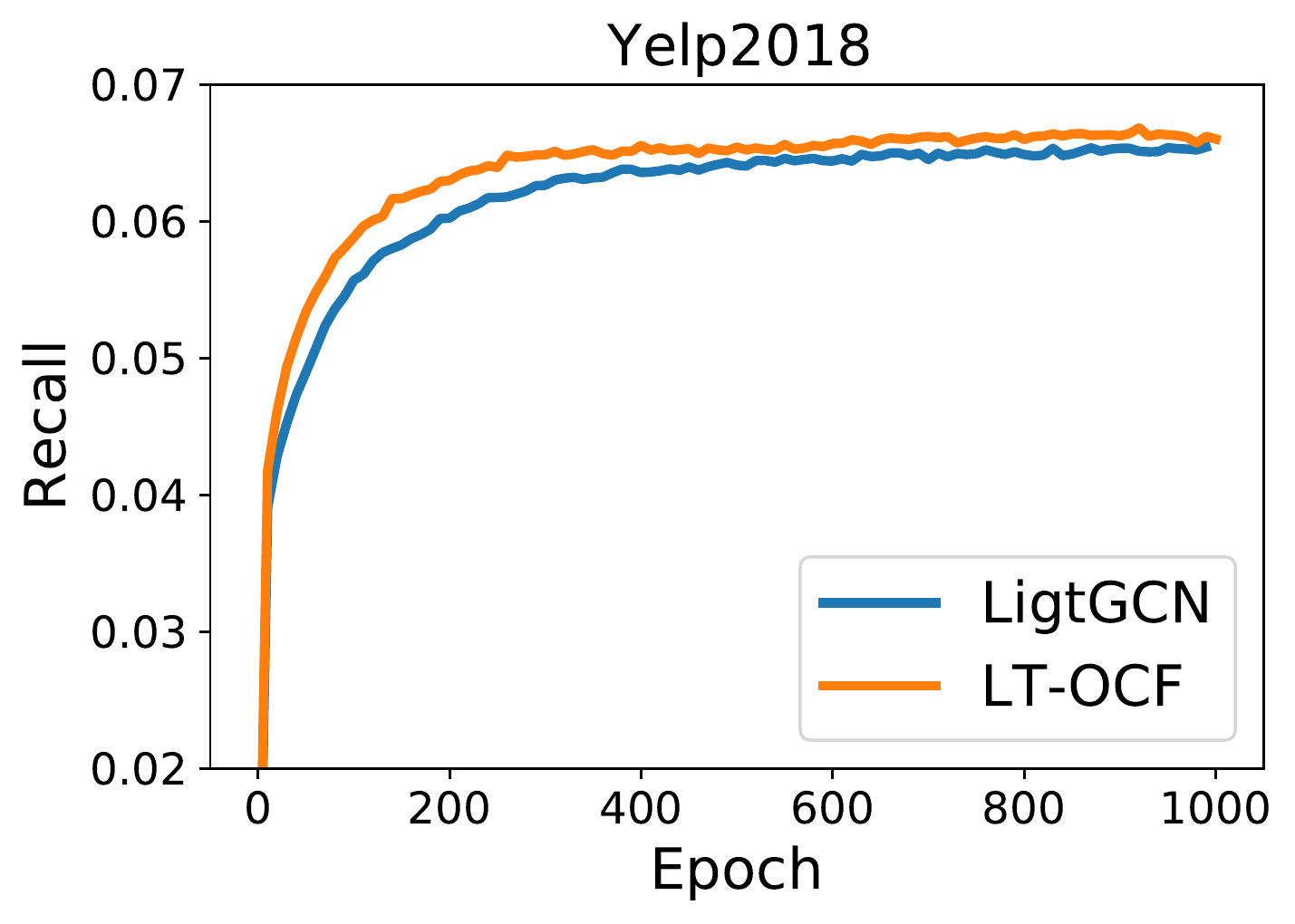}}
\subfigure[Training curve of NDCG]{\includegraphics[width=0.49\columnwidth]{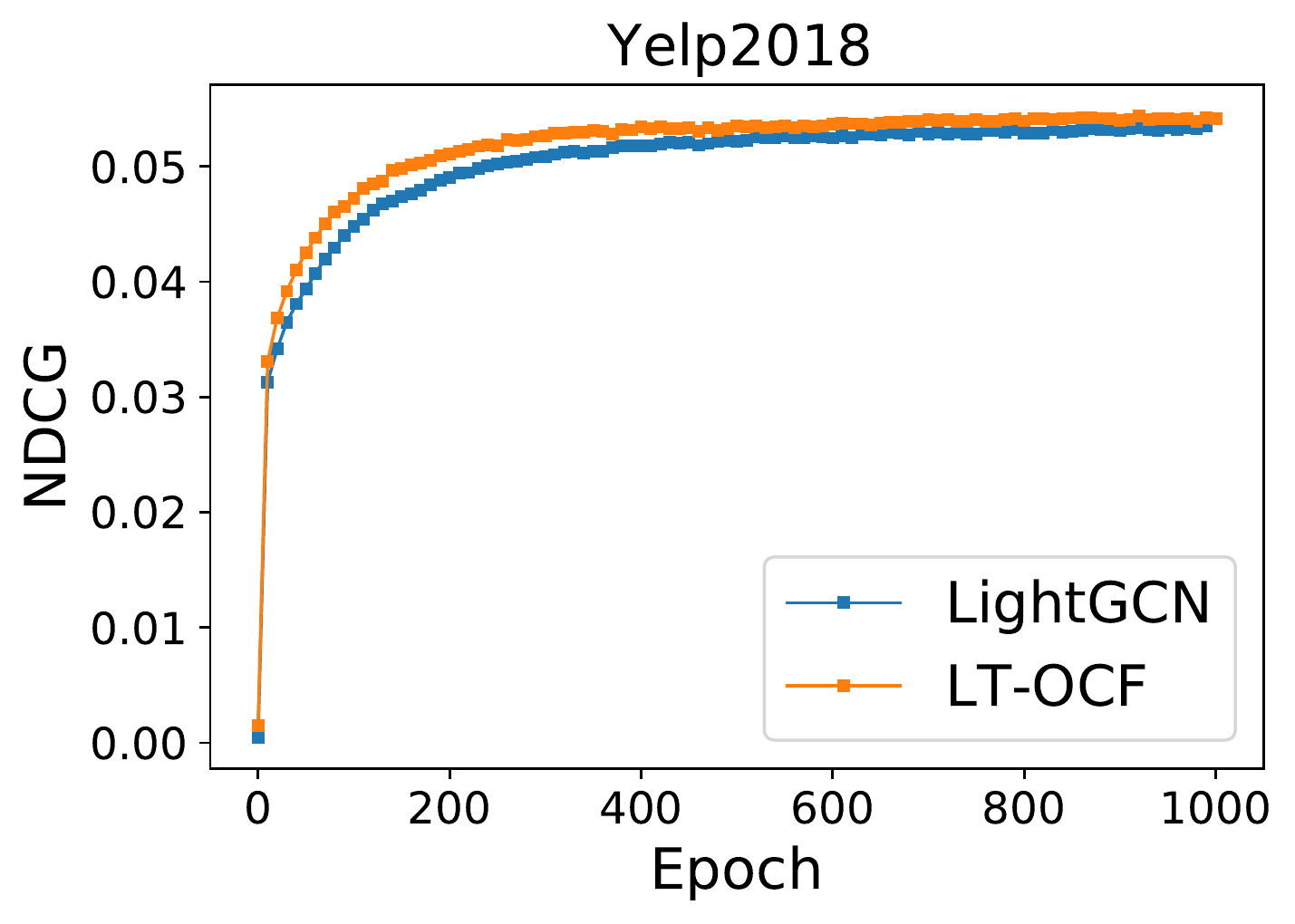}}
\subfigure[Training curve of $t_i$]{\includegraphics[width=0.49\columnwidth]{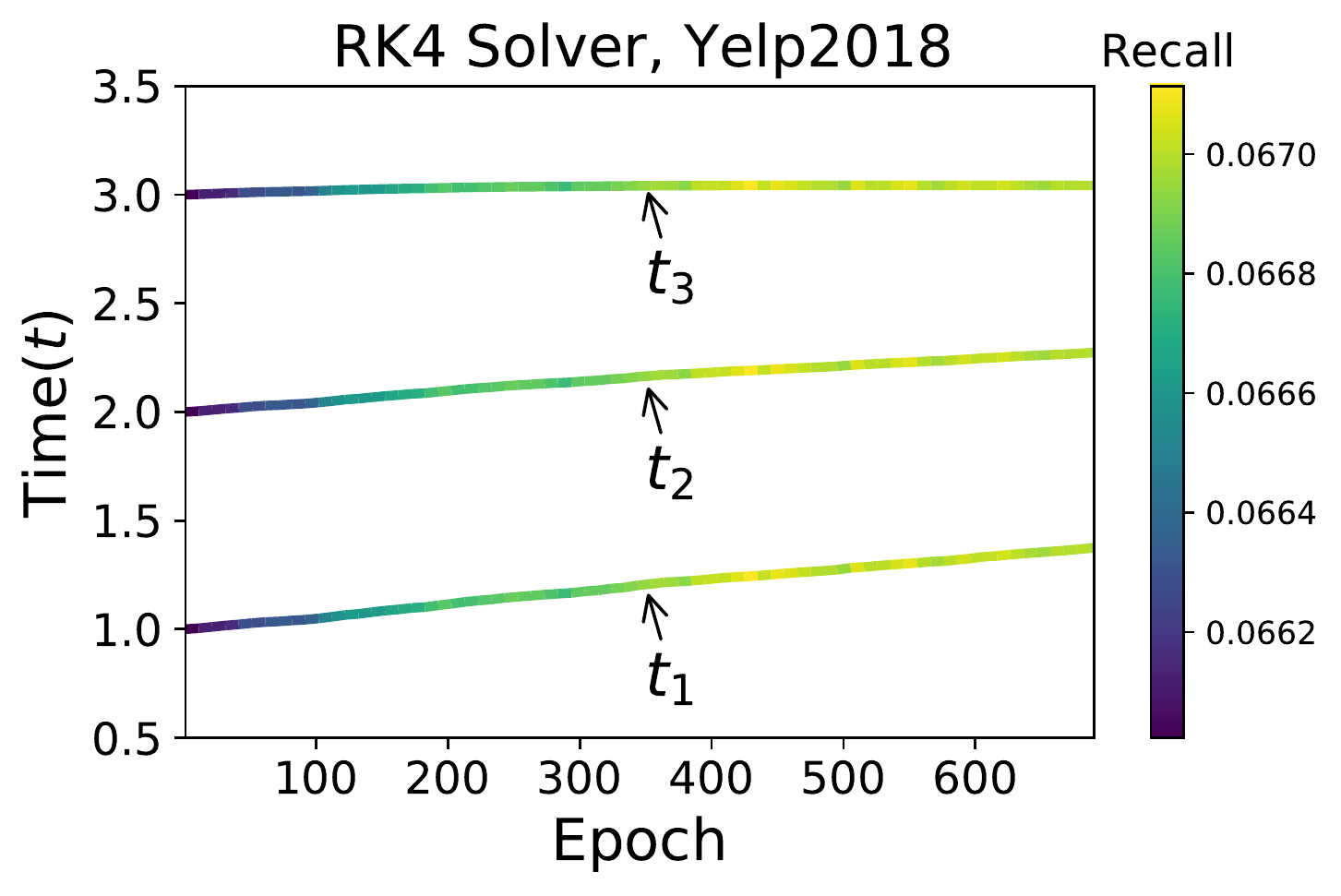}}
\caption{Training curves in Yelp2018. (a) The loss curves, (b) The training curves of Recall, (c) The training curves of NDCG, (d) The training curves of $t_1, t_2, t_3$ when $T=3$.} \label{fig:time2}
\end{figure}

\begin{figure}[t]
\centering
\subfigure[Training curve of loss]{\includegraphics[width=0.49\columnwidth]{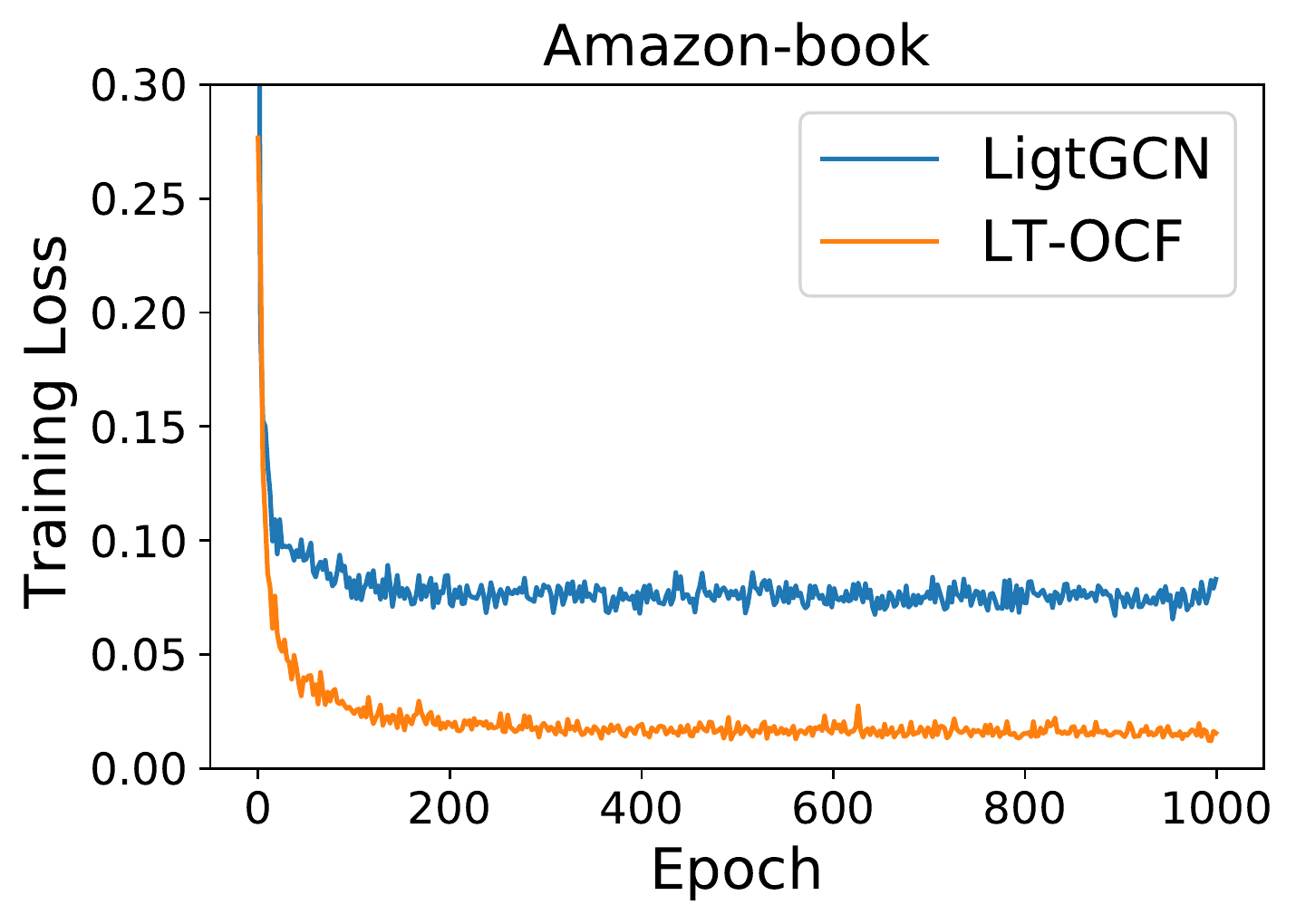}}
\subfigure[Training curve of recall]{\includegraphics[width=0.49\columnwidth]{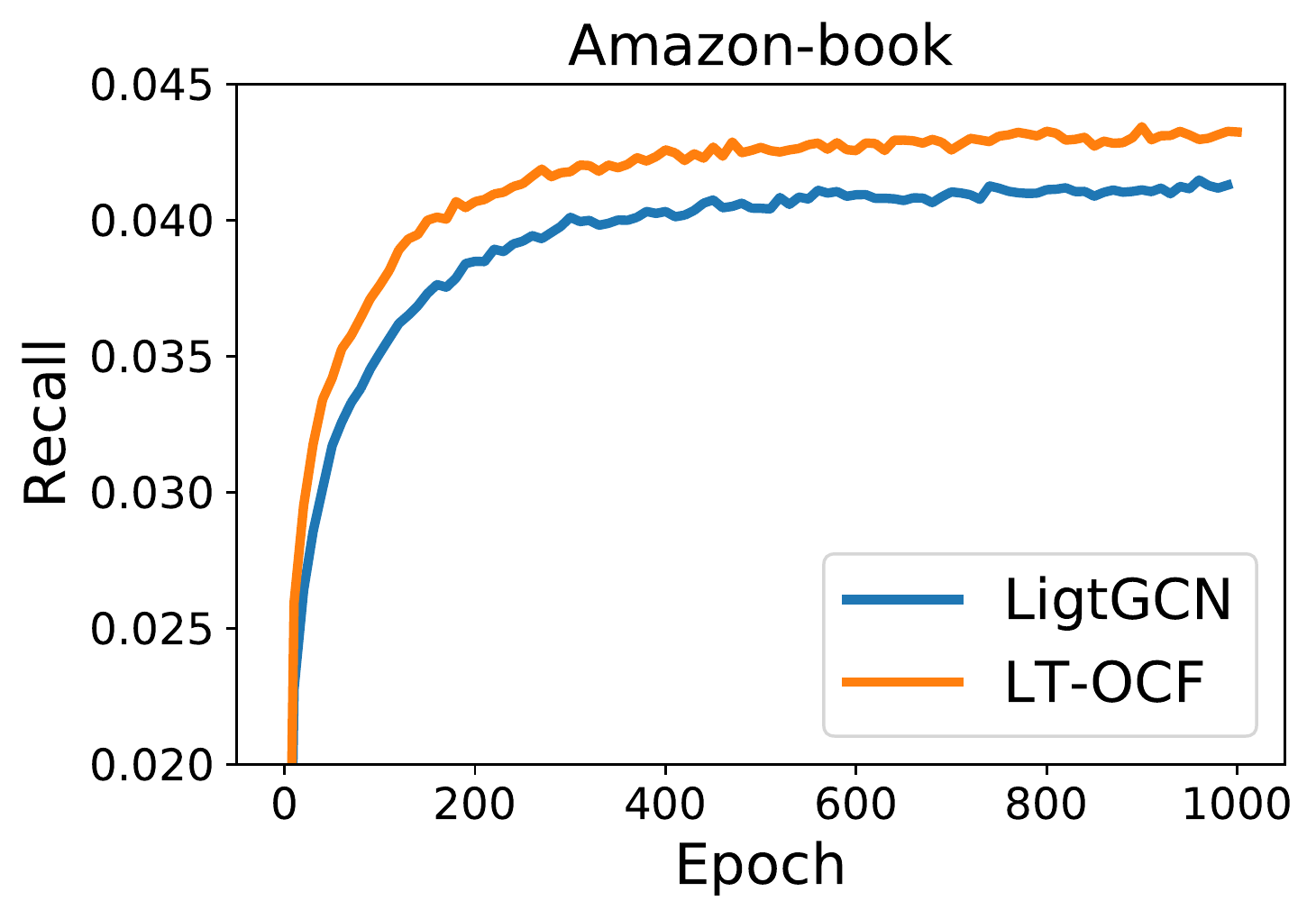}}
\subfigure[Training curve of NDCG]{\includegraphics[width=0.49\columnwidth]{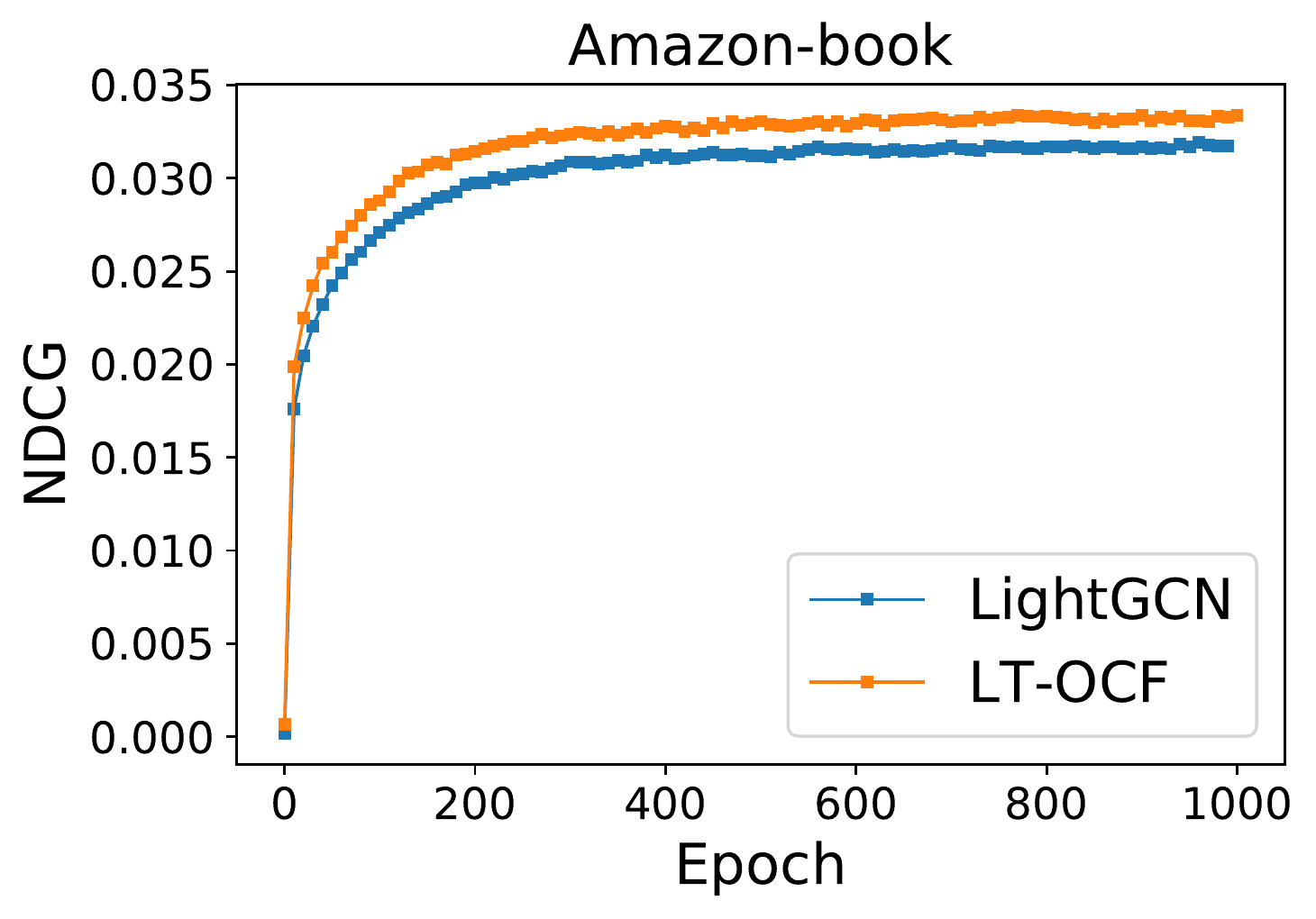}}
\subfigure[Training curve of $t_i$]{\includegraphics[width=0.49\columnwidth]{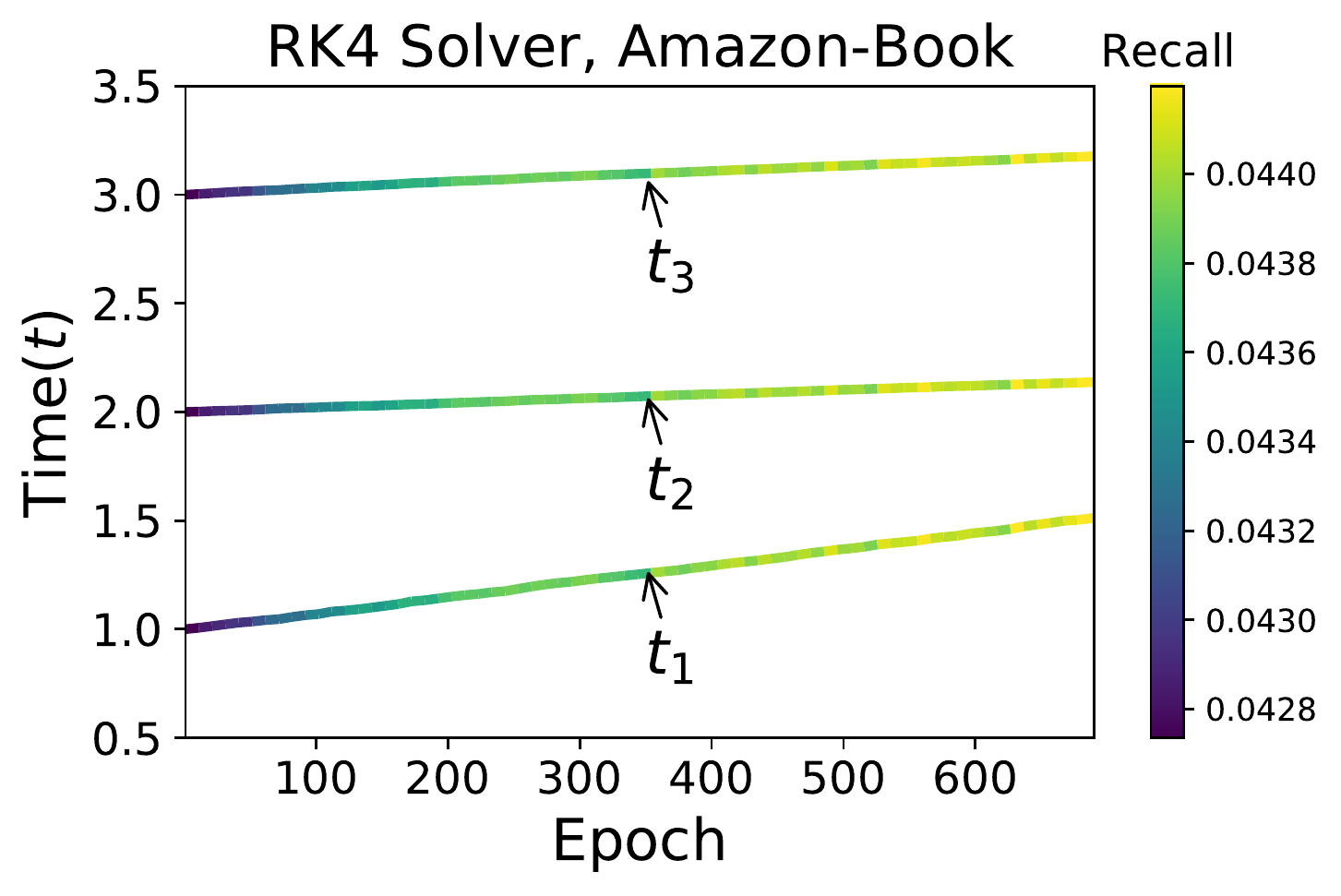}}
\caption{Training curves in Amazon-Book. (a) The loss curves, (b) The training curves of Recall, (c) The training curves of NDCG, (d) The training curves of $t_1, t_2, t_3$ when $T=3$.} \label{fig:time3}
\end{figure}

In Figure~\ref{fig:time1}, we compare the training curve of LightGCN and LT-OCF in Gowalla. In general, our method provides a faster training speed in terms of the number of epochs than that of LightGCN. In Figure~\ref{fig:time1} (d), we show that $t_i$ becomes larger (with a little fluctuation) as training goes on. It is because our model prefers embeddings from deep layers when constructing the layer combination. $t_1$ is more actively trained and $t_3$ is not trained much. According to this training pattern, we can know that it is more important to have reliable early layers for the layer combination.

In Figures~\ref{fig:time2} and~\ref{fig:time3}, we show the results of the same experiment types for Yelp2018 and Amazon-Book. For Amazon-Book, LT-OCF shows remarkably smaller loss values than that of LightGCN as shown in Figure~\ref{fig:time3} (a). In Figures~\ref{fig:time2} (b,c) and~\ref{fig:time3} (b,c), our method shows faster training for recall and NDCG than LightGCN. In Figure~\ref{fig:time3} (d), $t_1$ is trained a lot more than other cases in Figures~\ref{fig:time1} (d) and~\ref{fig:time2} (d)

\subsection{Ablation and Sensitivity Studies}
\begin{wrapfigure}{l}{0.25\textwidth}
\vspace{-1em}
    \centering
    \includegraphics[width=0.5\columnwidth]{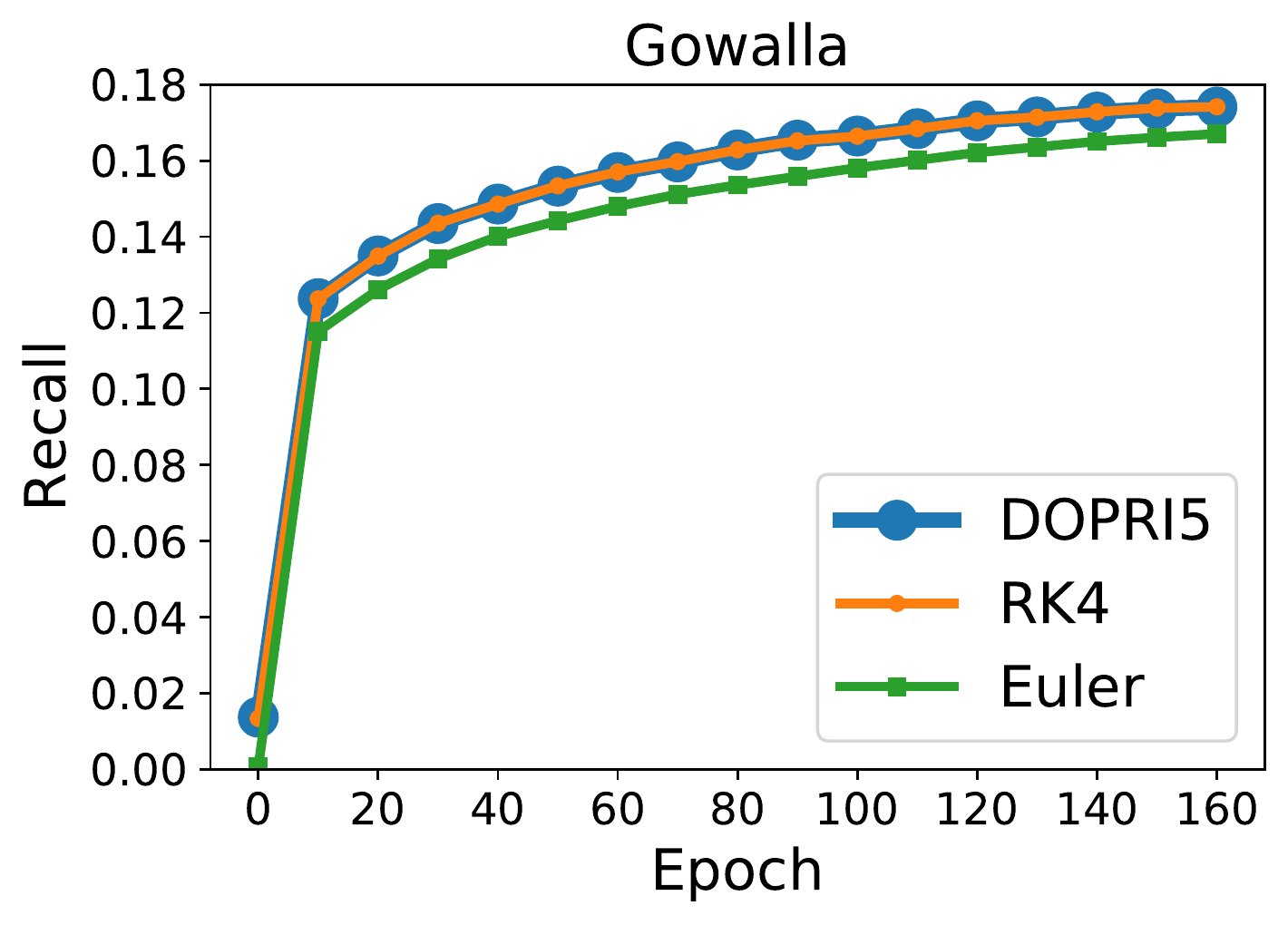}
    \caption{Various ODE solvers}
    \label{fig:solver}
\end{wrapfigure}

\subsubsection{Euler vs. RK4 vs. DOPRI} We first compare various ODE solvers. Figure~\ref{fig:solver} summarizes training curves of various ODE solvers. In general, DOPRI and RK4 are almost the same in terms of recall and NDCG while RK4 has 33\% smaller computation complexity. So, we use RK4 as our default solver. As mentioned earlier, RK4 shows better accuracy than that of the Euler method in solving general ODE problems and we observe the same result. For instance, our method ($K$=4, learning $t_i$) with the Euler method achieves a recall/NDCG of 0.1834/0.1548 vs. 0.1875/0.1574 with RK4 in Gowalla. For other datasets, we can observe similar patterns. RK4 consistently outperforms the Euler method. Adams-Moulto shows almost the same performance as that of RK4. However, it is an implicit ODE solver that requires more computation than other explicit methods, e.g., RK4.

\subsubsection{Sensitivity on $T$.} 
\begin{figure}[t]
\centering
\subfigure[Recall in Gowalla]{\includegraphics[width=0.49\columnwidth]{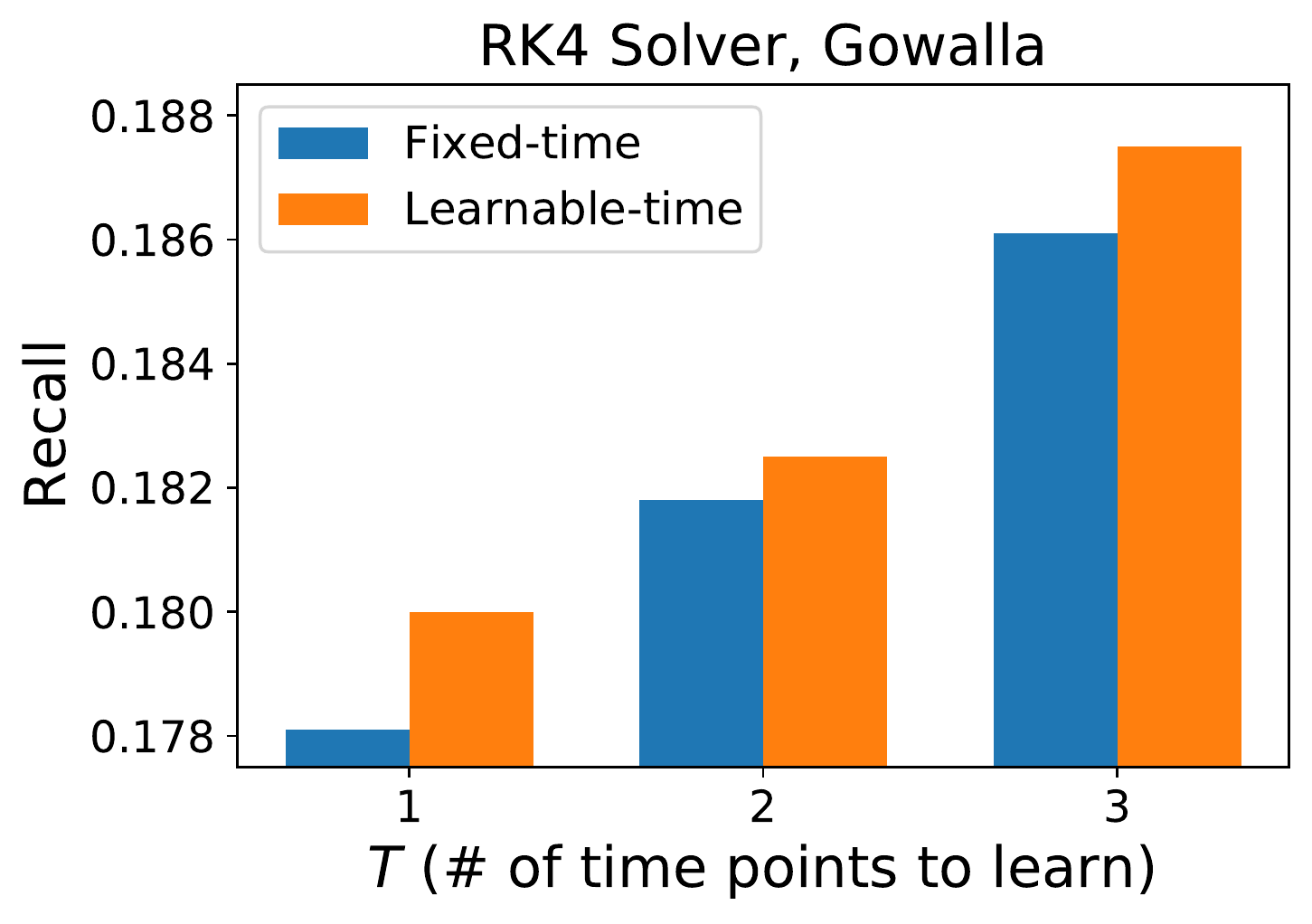}}
\subfigure[NDCG in Gowalla]{\includegraphics[width=0.49\columnwidth]{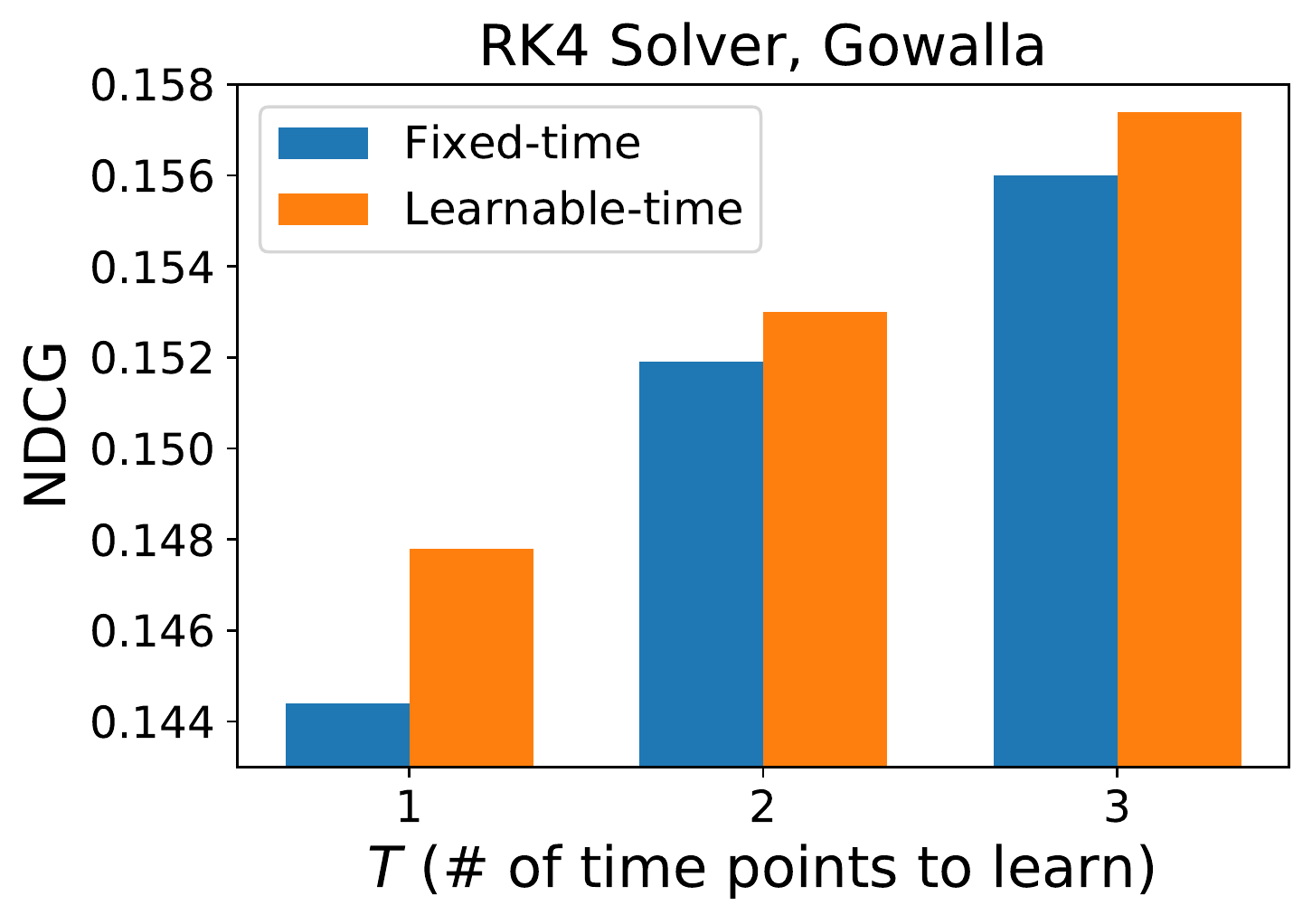}}
\subfigure[Recall in Yelp2018]{\includegraphics[width=0.49\columnwidth]{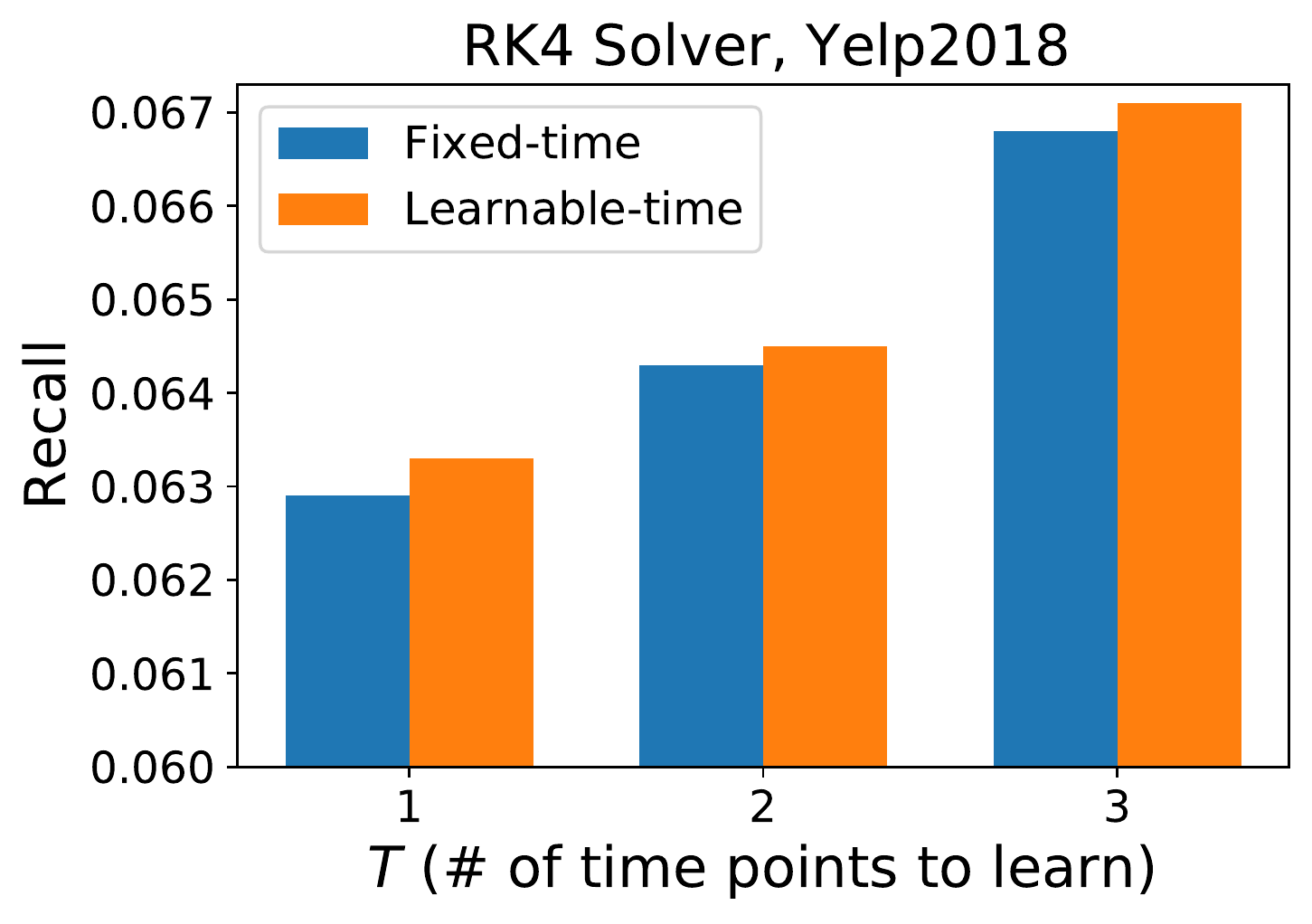}}
\subfigure[NDCG in Yelp2018]{\includegraphics[width=0.49\columnwidth]{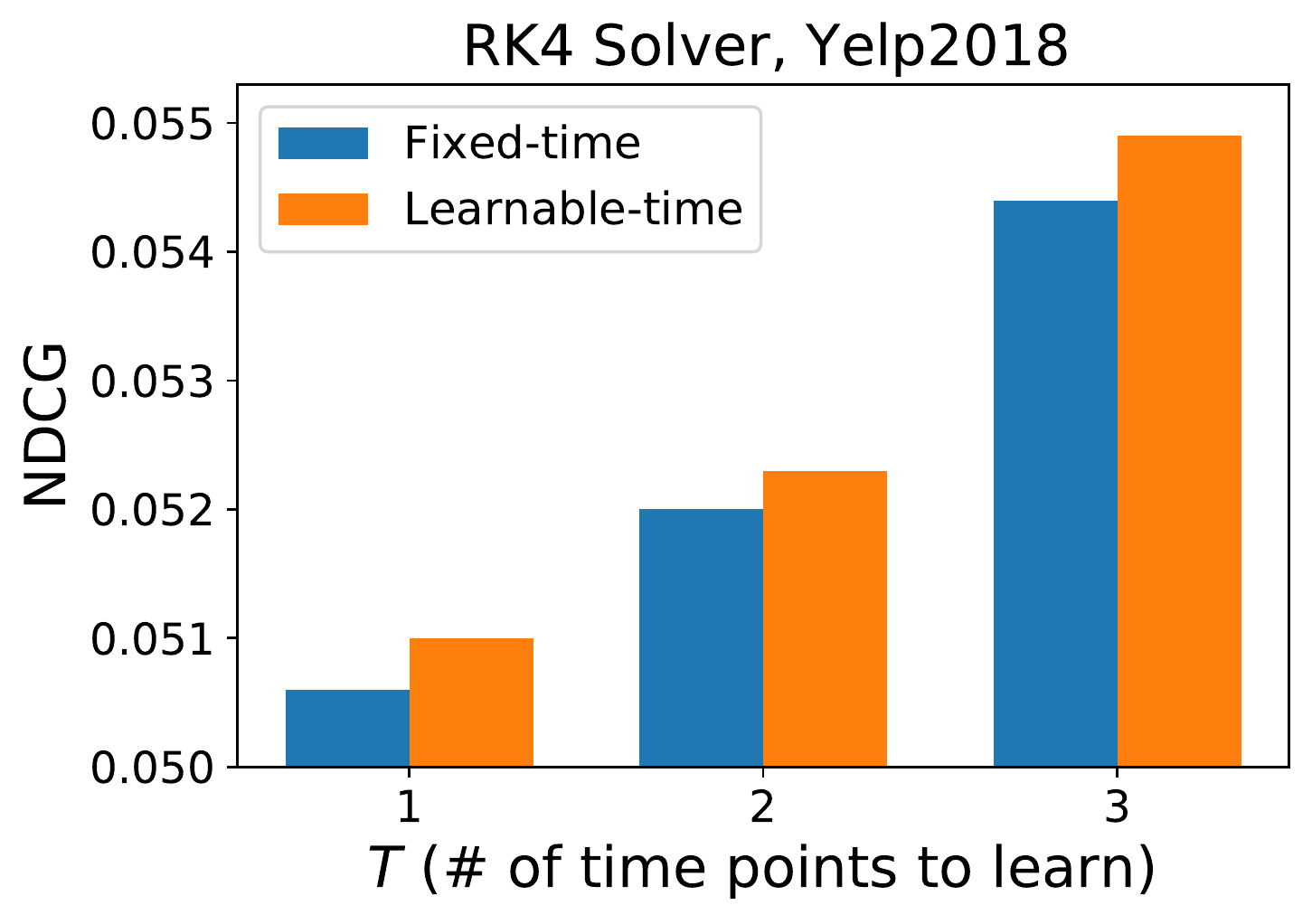}}
\caption{Performance comparison by varying $T$} \label{fig:tnum}
\end{figure}

By varying $T$, we also investigate how the model accuracy changes. The detailed results are in Figure~\ref{fig:tnum}. One point that is worth mentioning is that the fixed-time is sometimes more vulnerable to small $T$ than the learnable-time. In other words, the recall/NDCG gap between the fixed and the learnable-time at $T=1$ is larger than that in $T=2,3$ for Gowalla. In general, the recall increases as we increase $T$ but it is stabilized after $T=3$. Therefore, our best setting for $T$ is 3 in our experiments, considering computational efficiency. We can observe similar patterns in Amazon-Book as well.

\subsubsection{Sensitivity on $K$.} By varying $K$, we investigate how the model accuracy changes in Figure~\ref{fig:knum}. Our best results are all made with $K=4$. As decreasing $K$, we observe that performance also decreases. For instance, our method (RK4, learning $t_i$) achieves a recall/NDCG of 0.1833/0.1545 with $K=3$ and a recall/NDCG of 0.1823/0.1543 with $K=2$ in Gowalla. Similar patterns are observed in other two datasets.

\subsubsection{Fixed vs. Learnable-time.} Without learning $t_i$, we fix $t_i = \frac{K}{T+1}i$ and evaluate its accuracy. The learnable-time is one of the key concepts in our work. Without learning $t_i$, our method ($K$=4, RK4) still outperforms LightGCN in many cases but is consistently worse than our method with learning $t_i$. For instance, our method without learning $t_i$ achieves a recall/NDCG of 0.1859/0.1558 vs. a recall/NDCG of 0.1875/0.1574 by our method with learning $t_i$ in Gowalla. Similar patterns are observed in other two datasets.

\begin{figure}[t]
\centering
\subfigure[Recall in Gowalla]{\includegraphics[width=0.49\columnwidth]{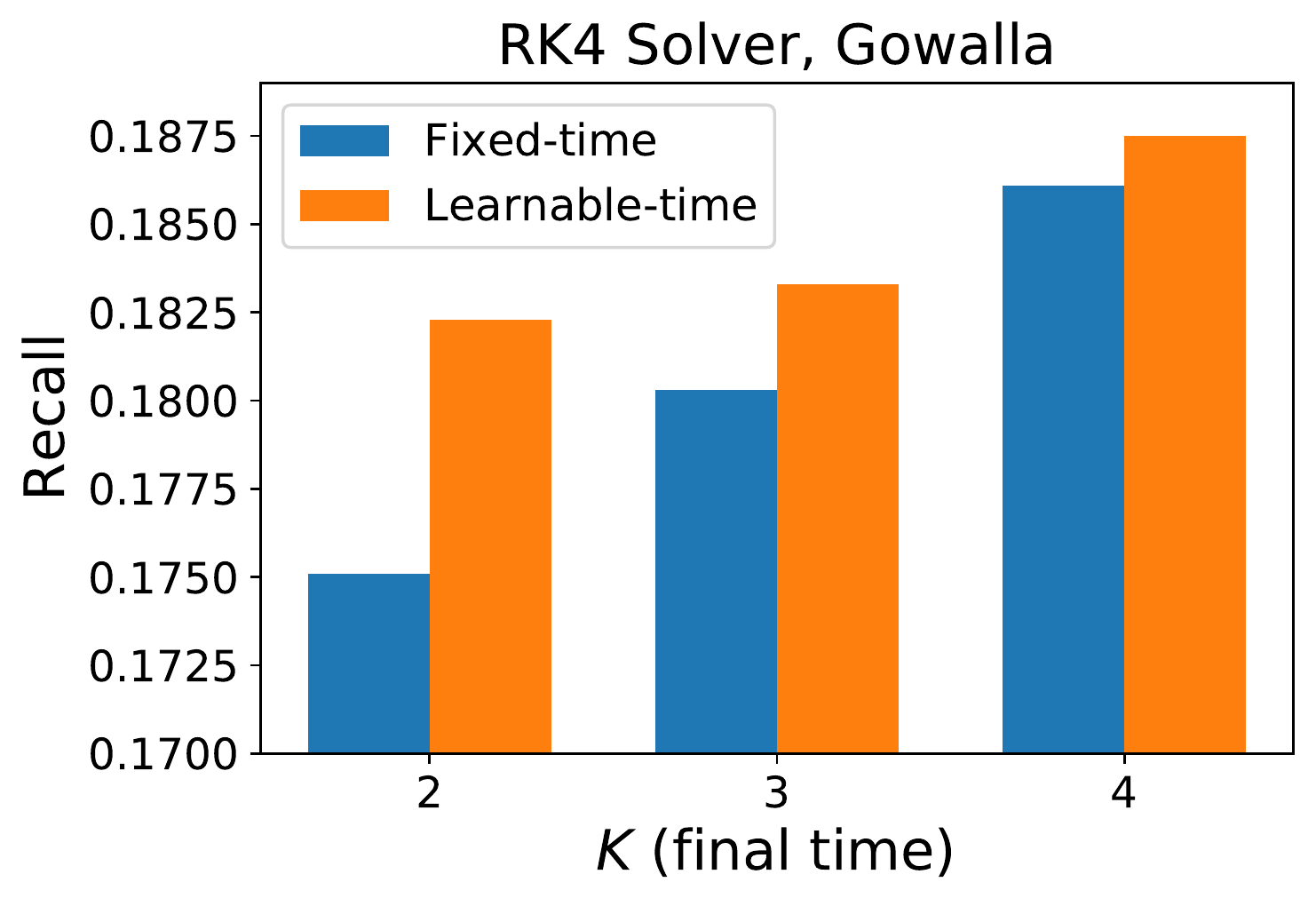}}
\subfigure[NDCG in Gowalla]{\includegraphics[width=0.49\columnwidth]{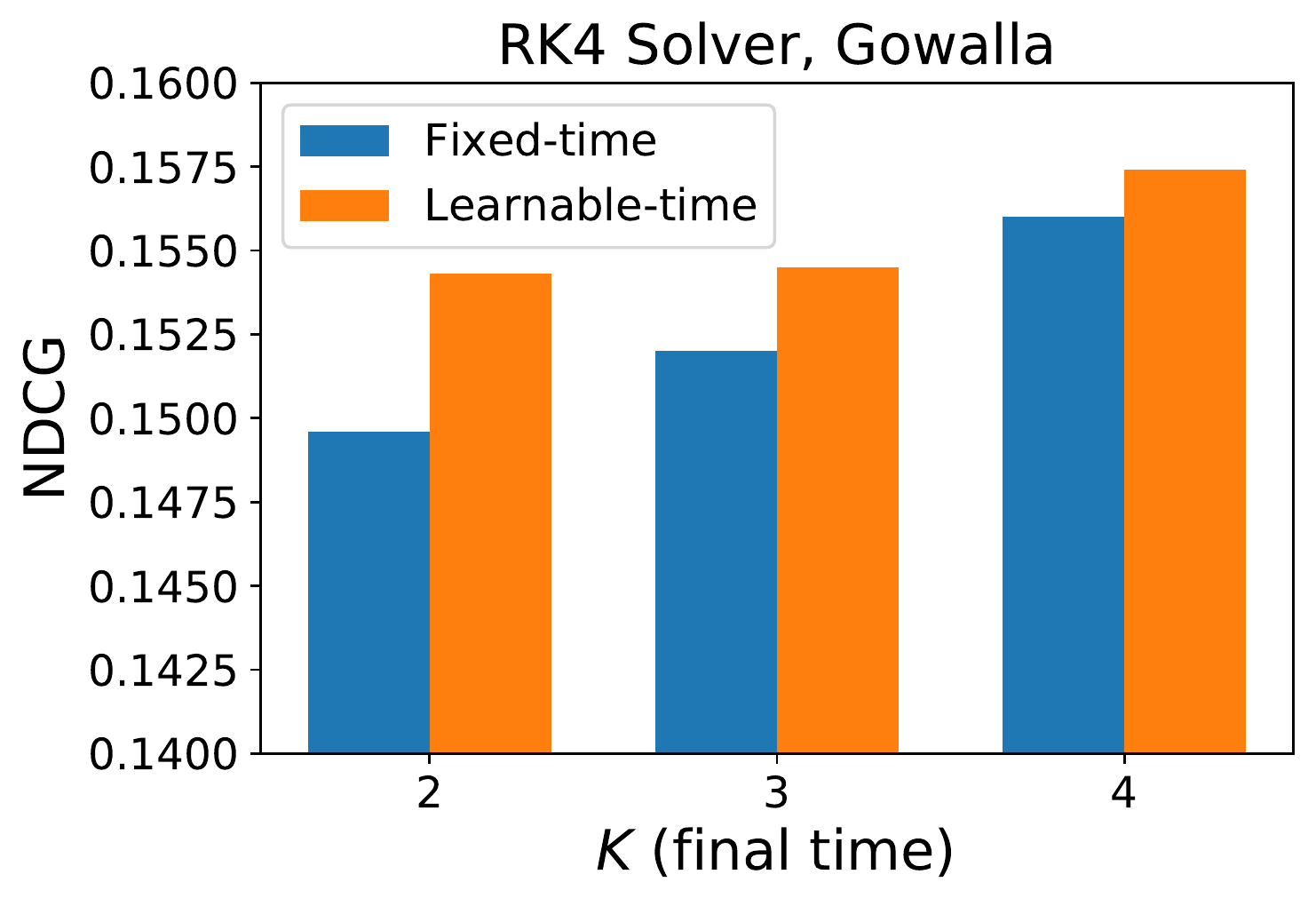}}
\subfigure[Recall in Yelp2018]{\includegraphics[width=0.49\columnwidth]{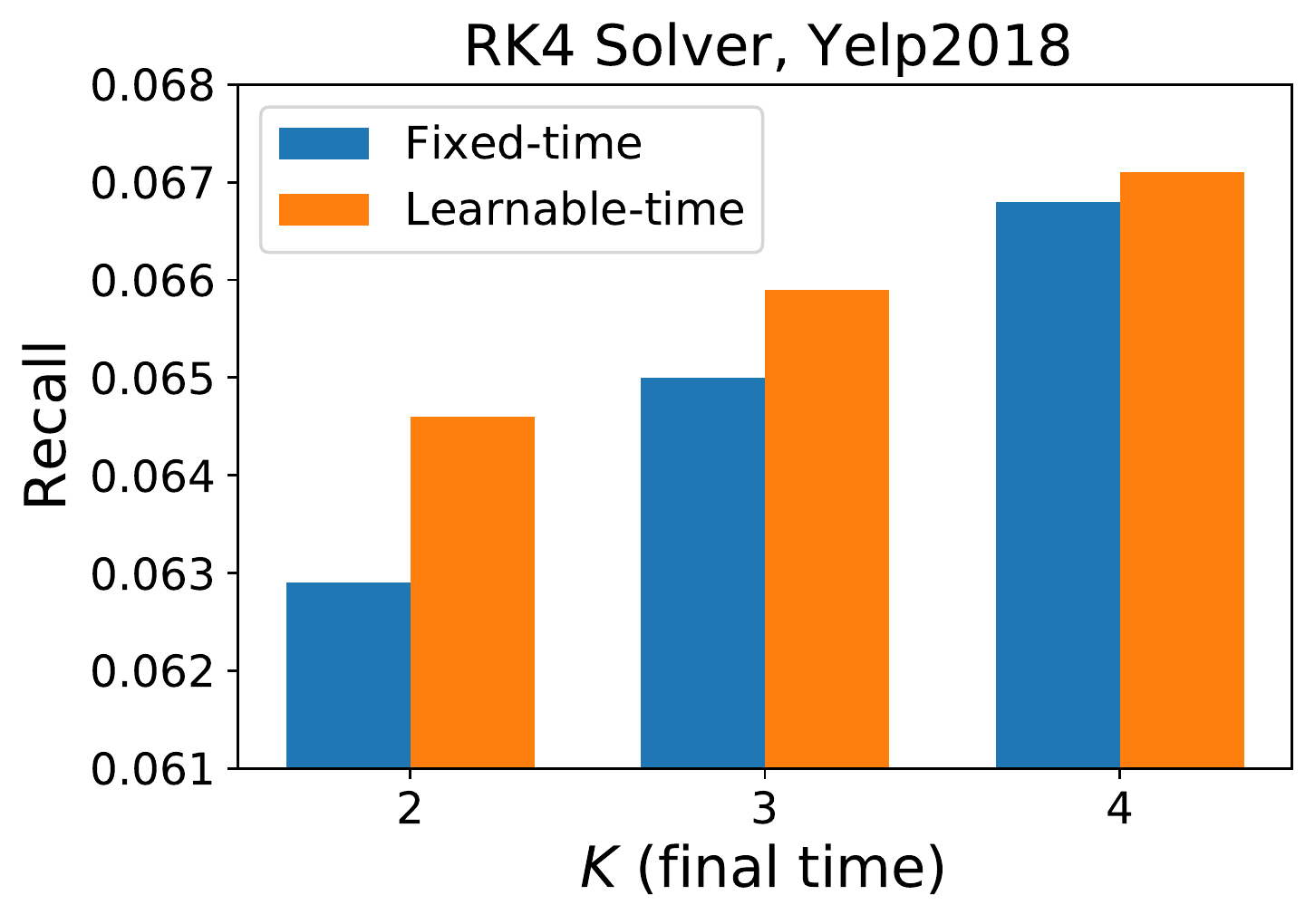}}
\subfigure[NDCG in Yelp2018]{\includegraphics[width=0.49\columnwidth]{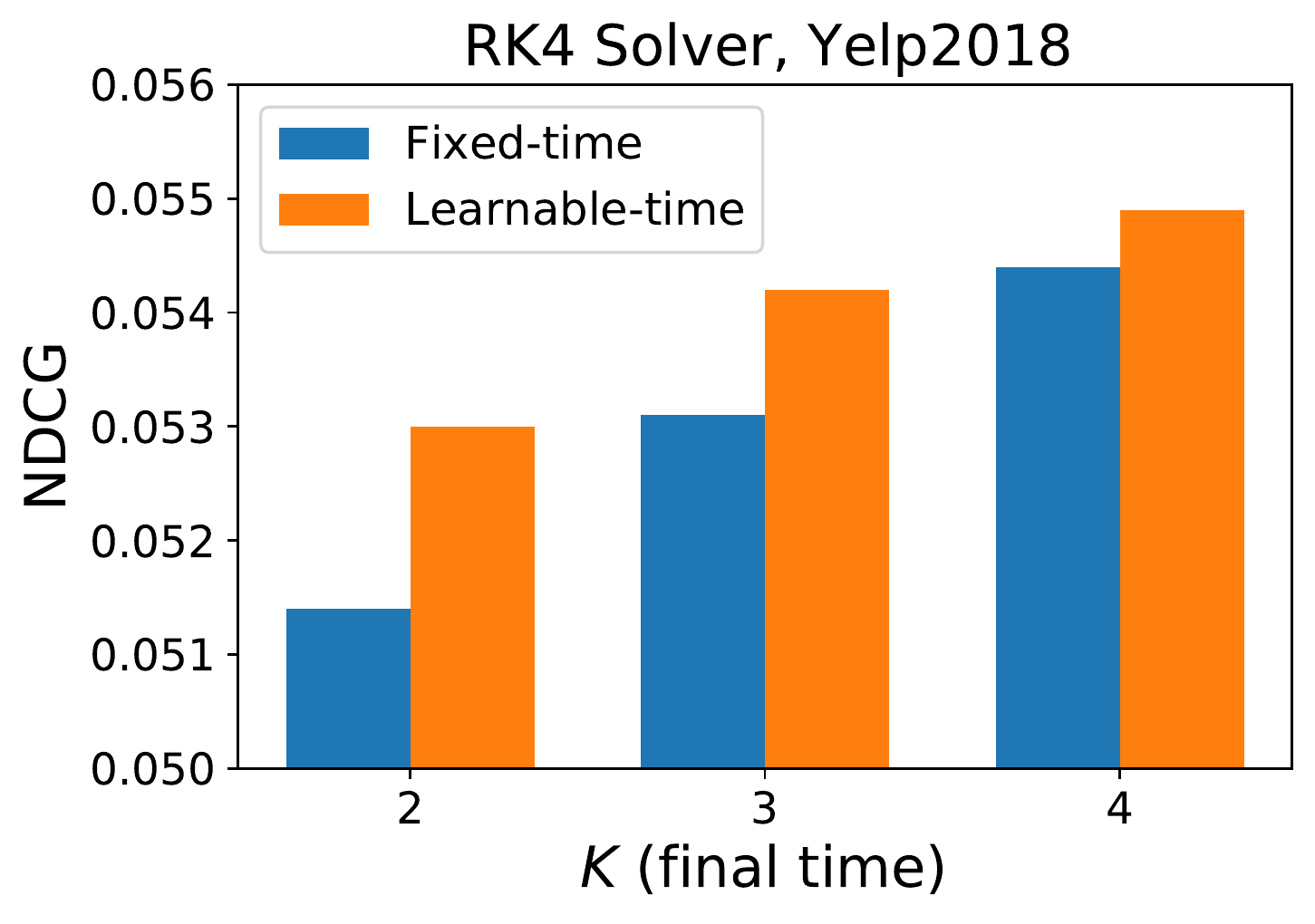}}
\caption{Performance comparison by varying $K$} \label{fig:knum}
\end{figure}

In particular, a combination of $K=2$ and fixed-time shows poor performance in all cases of Figure~\ref{fig:knum}. However, $K=2$ with learning-time surprisingly shows much improvement over it, which shows the efficacy of our proposed learnable-time concept.

\subsection{Runtime Analyses}

\begin{figure}[t]
\centering
\subfigure[Recall vs. training time]{\includegraphics[width=0.49\columnwidth]{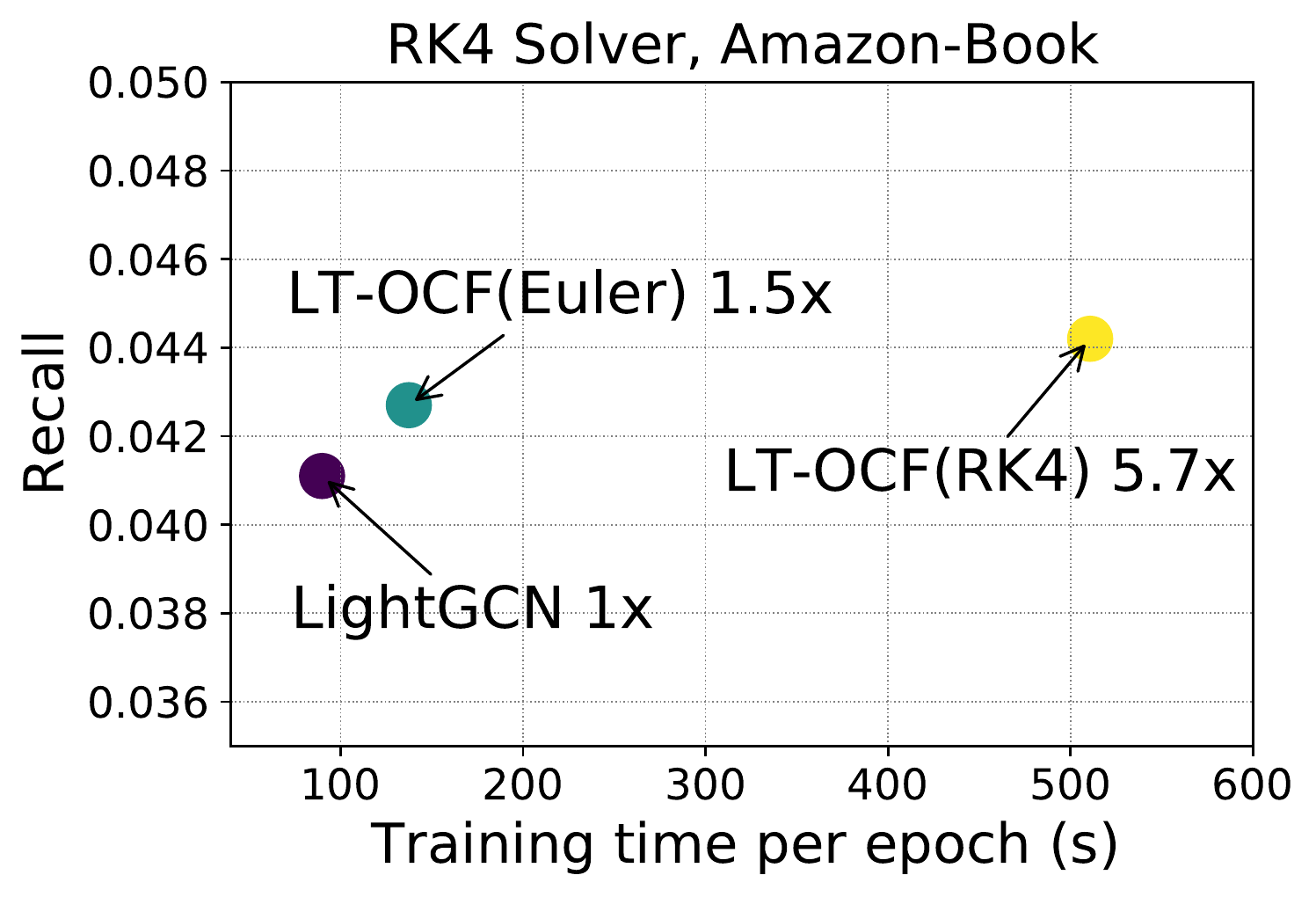}}
\subfigure[Recall vs. inference time]{\includegraphics[width=0.49\columnwidth]{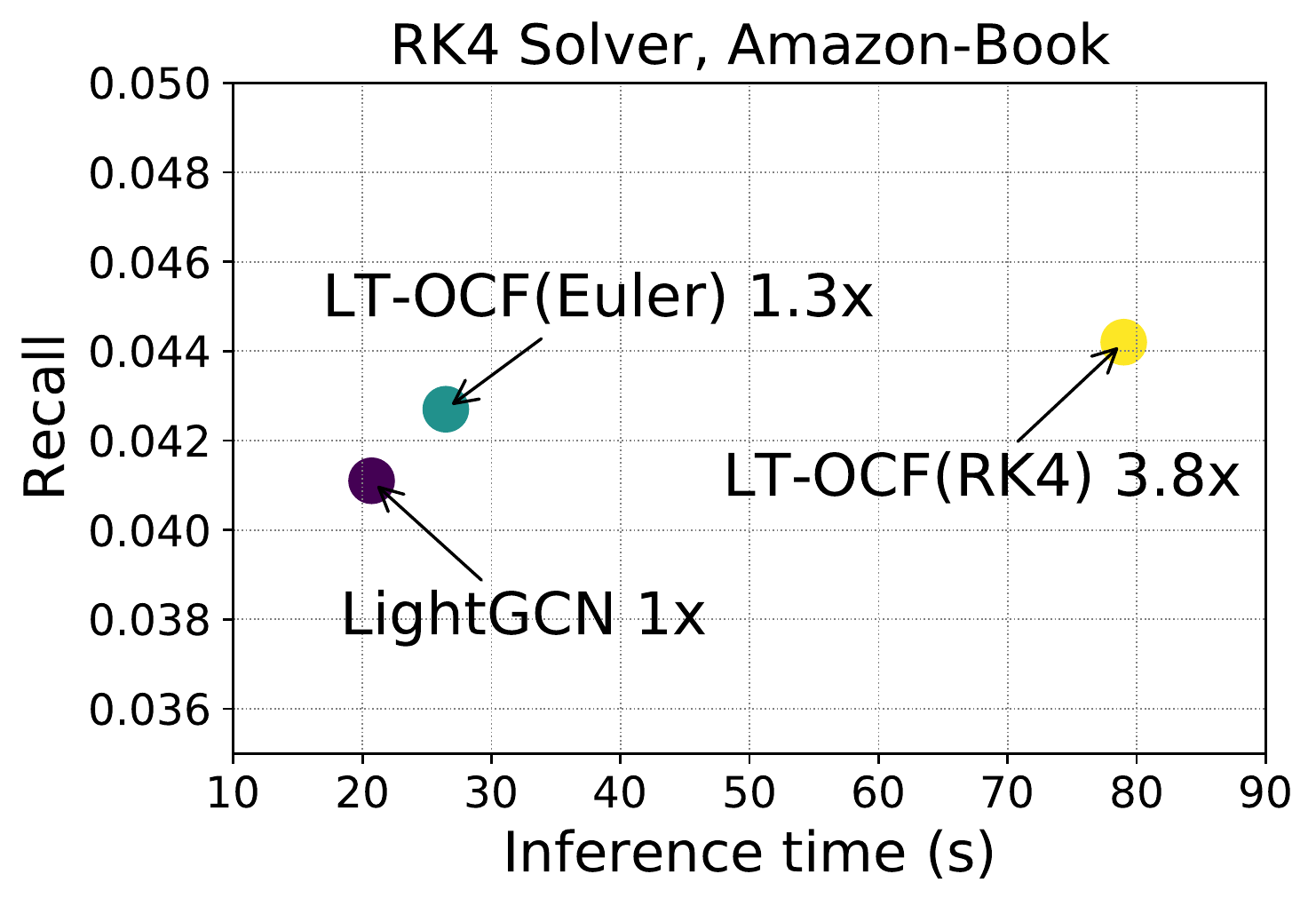}}

\caption{Training and inference time in Amazon-Book} \label{fig:runtime}
\end{figure}

We solve ODE problems in our method and as a result, need longer time to train and infer than other methods. We use Amazon-Book, the largest and the most suitable dataset for this subsection, to analyze the runtime of various algorithms. Among many baselines in Table~\ref{tbl:acc}, we mainly compare with LightGCN since it shows state-of-the-art accuracy and has a smaller complexity than other non-linear methods. As shown in Fig.~\ref{fig:runtime}, LightGCN is the fastest method for both training and testing. However, our method with the Euler method provides better recall scores in $30$ to $50$\% longer time. Even though RK4 shows the best accuracy, one can choose the Euler method to reduce the time complexity. In this regard, our proposed method is a versatile algorithm for CF, which provides a good trade-off between complexity and accuracy. Our method shows similar runtime patterns in other datasets as well. 

\section{Discussions on Linear vs. Dense}\label{sec:dis}
We revisit recent debates on figuring out the best GCN architecture for CF. It was recently reported that linear layers with a layer combination work well. However, we found that dense layers with a layer combination are better. As reported in the previous section, our best accuracy was all achieved by RK4, which internally constructs connections similar to DenseNet or FractalNet as described in Table~\ref{tbl:ode}~\cite{zhu2019convolutional,Larsson2017FractalNetUN,pmlr-v80-lu18d}. This is well aligned with the observation in ODEs that the explicit Euler method is inferior to RK4 in solve integral problems. We conjecture that dense connections are also optimal for non-ODE-based CF methods. We leave this as an open question.

\section{Conclusions}
We tackled the problem of learnable-time ODE-based CF. Our method fundamentally differs from other methods in that we interpret the user and product embedding learning process of CF as dual co-evolving ODEs.

Owing to the continuous nature of time variable $t$ in NODEs, we propose to train a set of time points $\{t_1, t_2, \cdots, t_T\}$, where we extract embedding vectors to construct a layer combination architecture. Our carefully designed training method guarantees a good solution by the well-posed nature of our formulation.

With the  benchmark datasets, our method, LT-OCF, consistently outperforms all state-of-the-art methods in all cases. We also showed that our method can be trained faster than other methods.

One more key contribution in this paper is that we revealed that dense connections, which are created by RK4, are the best option for our method and leave it as an open question to apply dense connections to other CF methods. We hope that this discovery will inspire forthcoming research works.

\begin{acks}
Noseong Park is the corresponding author. This work was supported by the Institute of Information \& Communications Technology Planning \& Evaluation (IITP) grant funded by the Korea government (MSIT) (No. 2020-0-01361, Artificial Intelligence Graduate School Program (Yonsei University)).
\end{acks}

\clearpage

\bibliographystyle{ACM-Reference-Format}
\bibliography{KDD21}

\end{document}